\documentclass{WileyMSP-template}
\usepackage{multicol}
\usepackage{setspace}
\usepackage{amsmath}
\usepackage{lineno}
\usepackage{caption}  
\usepackage{braket}
\usepackage{hyperref}
\usepackage{textcomp}
\justifying
\hypersetup{colorlinks,linkcolor={blue},citecolor={blue},urlcolor={red}}  

\nolinenumbers
\renewcommand{\subsectionmark}[1]{}
\begin{document}
\pagestyle{fancy}
\rhead{\includegraphics[width=2.5cm]{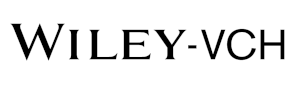}}

\title{Stark-tunable O‑band single-photon sources based on deterministically fabricated quantum dot - circular Bragg gratings on silicon}
\maketitle

\vspace*{2mm}

\author{Sarthak Tripathi$^{1}$},
\author{Kartik Gaur$^{1}$},
\author{Priyabrata Mudi$^{1}$},
\author{Peter Ludewig$^{2,3}$},
\author{Alexander Kosarev$^{1}$},
\author{Kerstin Volz$^{2,3}$},
\author{Imad Limame$^{1}$},
\author{and Stephan Reitzenstein$^{1}$*}

\vspace*{2mm}

\begin{affiliations}

{$^1$Institut für Physik und Astronomie, Technische Universit\"at Berlin, Hardenbergstrasse 36, 10623 Berlin, Germany}

{$^2$mar.quest - Marburg Center for Quantum Materials and Sustainable Technologies, Philipps-Universität Marburg, Hans Meerwein Str. 6, 35032 Marburg, Germany}

{$^3$Department of Physics, Philipps-Universität Marburg, Structure and Technology Research Lab, Hans Meerwein Str.~6, 35032 Marburg, Germany}

\end{affiliations}

*Corresponding author: stephan.reitzenstein@physik.tu-berlin.de


\begin{abstract} 
Semiconductor quantum dots (QDs) provide outstanding quantum-optical performance, which is highly interesting for applications in quantum information technology. However, the simultaneous realization of wide-range electrical tunability, efficient photon extraction, elevated-temperature operation, and monolithic integration on silicon in combination with telecom-wavelength compatibility remains a major unresolved challenge. Here, we demonstrate electrically contacted circular Bragg grating (eCBG) resonators incorporating InGaAs QDs directly grown on silicon, enabling bright single-photon emission in the telecom O-band. Deterministic electron-beam lithography and a ridge-based vertical p-i-n diode architecture allow precise device integration and electrical control of individual emitters. The QD-eCBGs exhibit a quantum-confined Stark shift of approximately 16~nm~(11~meV) at 4~K, which is the record value for a QD embedded in a nanophotonic structure at telecom wavelengths. This feature is combined with a photon extraction efficiency of $(21.7 \pm 3.0)\%$ into the first lens, while preserving radiative dynamics and excellent single-photon purity, with $g^{(2)}(0)= 0.0078\pm0.0012$ below saturation and $g^{(2)}(0)=0.0183\pm0.0021$ at saturation under pulsed excitation. Robust antibunching persists up to 77~K, yielding $g^{(2)}(0)=0.0663\pm0.0056$, demonstrating operation within the temperature range accessible to liquid-nitrogen and compact, user-friendly Stirling cryocoolers. Moreover, spatially separated QD-eCBGs can be electrically fine-tuned into spectral resonance without degrading photon statistics. With this, our results establish a silicon-compatible, electrically addressable telecom O-band quantum light platform combining wide-range spectral tunability, high single-photon purity, and elevated-temperature operation, providing a scalable route toward practical photonic quantum networks.
\end{abstract}


\section{Introduction}\label{sec1}

Deterministic sources of non-classical light constitute key building blocks for quantum communication and quantum information processing~\cite{Su2020}. Among solid-state platforms, semiconductor quantum dots (QDs) have emerged as one of the most mature and versatile single-photon emitter systems, enabling on-demand single-photon generation, high photon indistinguishability, and the generation of entangled photon pairs~\cite{Muller2014, Ding2016, Huber2017, Senellart2017, Liu2019}. Together with their compatibility with advanced semiconductor nanofabrication techniques, these properties make QDs a leading platform for scalable integrated quantum photonic circuits and devices~\cite{Uppu_2020_SA, Heindel2023, Wang2025}.

While the highest performance of QD-based single-photon sources (SPS) has traditionally been realized in the near-infrared (NIR) spectral range~\cite{Ding2016, Somaschi2016, Senellart2017, Wang2019, Tomm2021}, the flexibility of III-V material systems enables extension toward telecommunication wavelengths~\cite{Lee2021, Kolatschek2021, Holewa2024, Hauser2026},  a regime of critical importance for long-distance quantum communication, where propagation losses in standard silica fibers are minimized~\cite{Gisin2002}. In particular, the telecom O-band near 1.3 $\mu$m provides near-zero chromatic dispersion, while the C-band near 1.55 $\mu$m offers minimal fiber attenuation, together defining the optimal spectral window for fiber-based quantum networks~\cite{Agarwal2021}.

One concept to achieve telecom compatibility is based on nonlinear frequency conversion of NIR QD emission~\cite{ZaskePRL2012, Singh2019}. Although this approach can preserve photon coherence, it requires additional nonlinear optical components and auxiliary laser systems and is ultimately limited by conversion efficiency and increased experimental complexity. These constraints motivate the development of QDs operating intrinsically at telecom wavelengths. Direct telecom emission from QDs has been achieved through advanced epitaxial and material engineering strategies. In the InAs/GaAs system, emission in the telecom O-band can be achieved via strain engineering~\cite{Nishi1999, HOSPODKOVA2007, Paul_2015}, while the InAs/InP material system inherently accesses the C-band~\cite{Benyoucef_2013, Miyazawa_2016}. These advances, together with progress in nanofabrication, have enabled compact, fiber-compatible quantum light sources~\cite{Musial_2020, Gao_2022, Yang2024}, though practical deployment requires not only spectral compatibility but also integration with scalable photonic platforms.

Beyond telecom-wavelength compatibility, the choice of material platform plays a decisive role in determining the application relevance of quantum light sources. Silicon photonics offers an attractive foundation in this regard, combining established fabrication processes, high reproducibility, and compatibility with existing optical interconnect technology. Significant advances have been made in post-growth integration
techniques, such as wafer bonding and transfer printing for combining III-V heterostructures with silicon substrates~\cite{Katsumi2019, Vijayan2024}. Although the monolithic growth of III-V quantum emitters on silicon substrates has long been hindered by lattice mismatch and defect formation, recent advances in buffer-layer engineering and defect filtering layers have enabled the direct growth of optically active InGaAs QDs on silicon substrates, paving the way for silicon-integrated quantum photonic devices~\cite{Luxmoore2013, Tang2014, Limame2024}.

Of additional concern is the operating temperature of an SPS, which critically determines its practical viability. Unlike probabilistic room-temperature sources~\cite{Paesani2020}, InGaAs QDs offer deterministic on-demand single-photon generation, yet room-temperature single-photon emission at telecom wavelengths remains constrained because of increased phonon-induced dephasing and thermally activated carrier escape~\cite{Olbrich_2017, Holewa2020TelecomQD}. Still, user-friendly QD SPSs can be realized by using compact Stirling cryocoolers with a base temperature of about 30 K~\cite{Schlehahn_2015}, including sources operating in the telecom O-band ~\cite{Musial_2020}. 

Practical sources further demand nanophotonic structures with high, and preferably broadband, photon-extraction efficiency (PEE), a need directly addressed by circular Bragg grating (CBG) resonators, which combine broadband emission enhancement with directional vertical outcoupling in a compact geometry~\cite{Davanco_2011, Sapienza2015}. Such architectures are a key ingredient for scalable quantum photonic platforms, where reproducible device performance must be maintained across multiple nominally identical structures~\cite{Gaur2025Scalable}. Although site-controlled QD growth strategies show promise for improved spatial control and reproducibility through deterministic emitter positioning during epitaxy~\cite{Baier_2004, Schneider_2008, Gaur2025MQT, Limame2026SCQDs}, overcoming the intrinsic spectral variability of self-assembled QDs remains crucial for practical device implementation. Electrical control via the quantum-confined Stark effect (QCSE)~\cite{Mille1984, Finley_2004} provides a direct and reversible mechanism to tune the excitonic transition energy by modifying the electric field across the QD~\cite{Bennett2010}. This field-induced energy shift enables precise wavelength control without changing the excitation conditions or device temperature and offers a practical route to compensate spectral variations between individual emitters. Implementing such electrical tuning within nanophotonic structures while maintaining high PEE has long remained a significant challenge, one that electrically contacted CBGs (eCBGs) have recently begun to address by combining spectral tunability and efficient photon extraction within a single compact device architecture~\cite{Wijitpatima2024}.

In this article, we report on the deterministic integration of InGaAs QDs monolithically grown on silicon into eCBG resonators for single-photon emission in the telecom O-band. The underlying heterostructure, comprising a distributed Bragg reflector (DBR), an embedded QD layer, and a vertical p-i-n diode, is realized on a silicon substrate with a GaAs epi-ready surface enabled by an optimized GaP layer stack. Individual QD emitters are deterministically positioned within the central mesa of the eCBG resonator and electrically connected to the outer contact via a narrow conducting ridge, enabling simultaneous enhanced photon extraction and precise electrical control within a monolithic quantum device design. Through bias-dependent micro-photoluminescence ($\mu$PL) spectroscopy and photon-autocorrelation measurements, we demonstrate a record high Stark tuning range of approximately $16\,\mathrm{nm}$ ($11\,\mathrm{meV}$) in a nanophotonic structure and high single-photon purity with $g^{(2)}(0)$ as low as 0.0078 ± 0.0012 and persisting up to $77\,\mathrm{K}$ with $g^{(2)}(0) = 0.0663 \pm 0.0056$ at saturation. Crucially, the demonstrated tuning range enables two independent emitters to be brought into mutual spectral resonance without degrading their single-photon characteristics, establishing a key requirement for multi-emitter quantum photonic applications, such as quantum repeater networks based on entanglement distribution between remote quantum light sources~\cite{Azuma2023}. These results establish electrically tunable, silicon-compatible telecom O-band SPSs and represent an important step toward scalable integrated quantum photonic circuits and fiber-based quantum communication networks.

\section{Design optimization and fabrication}\label{sec1}

This section presents the realization of the electrically contacted QD-eCBG devices. We first discuss the epitaxial heterostructure and cavity design optimized for telecom O-band operation, followed by the eCBG device concept and numerical simulations, and finally the deterministic marker-based fabrication process enabling precise integration of individual QDs into the eCBG resonators.

\subsection{Epitaxial growth and layer design}\label{subsec1}

\begin{figure}[h]
 \centering
  \includegraphics[width=1.0\textwidth]{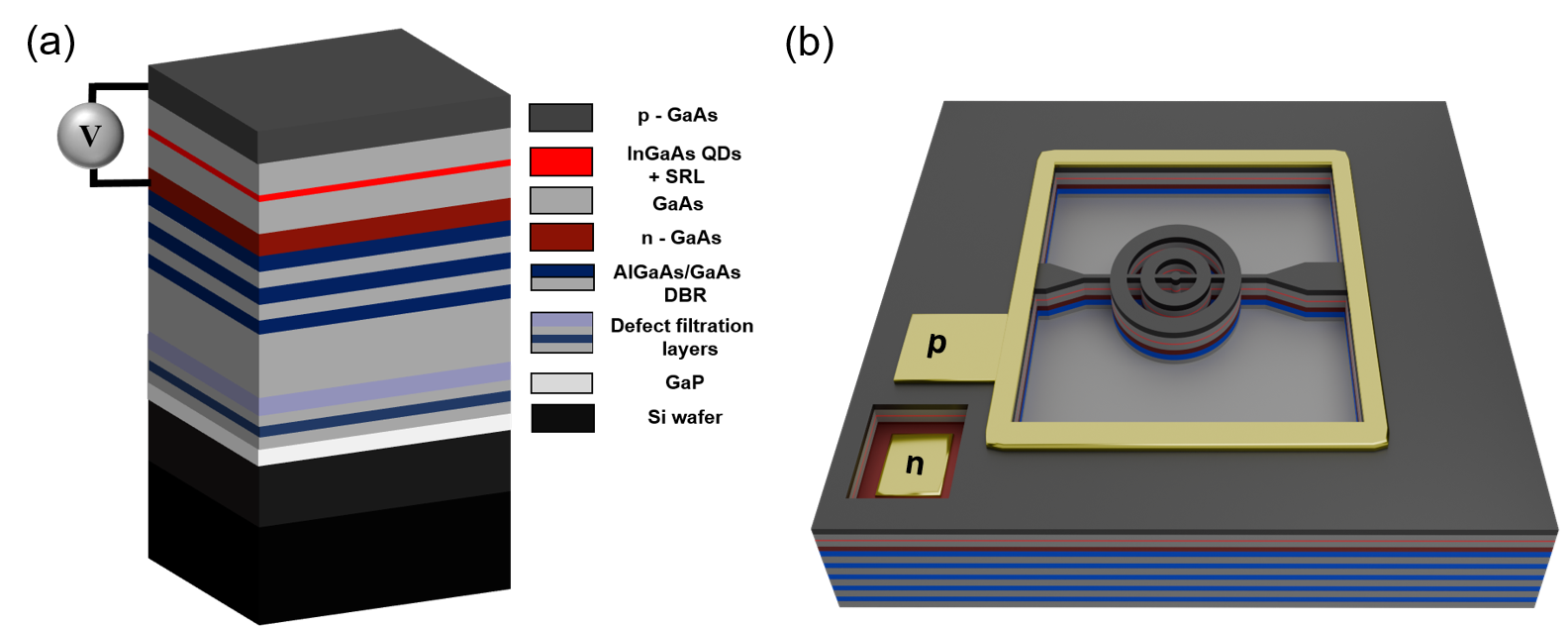}
  \caption
  {\textbf{Layer and device design of the QD-eCBG quantum light sources. (a)} Schematic of the monolithically grown III-V heterostructure on silicon \textbf{(b)} Three-dimensional illustration of the electrically contacted circular Bragg grating, showing the central mesa, concentric Bragg rings, and the electrical ridge required for biasing.}
  \label{fig:1}
\end{figure}

The developed quantum light sources are based on the direct growth of a III-V heterostructure on exact-oriented Si(001) substrate, schematically shown in Fig.~\ref{fig:1}(a), which is realized via a two-step metal-organic chemical vapor deposition (MOCVD) process. The first epitaxial stage is carried out in an AIXTRON Crius R CCS twin-reactor system optimized for III-V growth on silicon. Growth commenced with a 5~$\mu$m thick Si:P buffer layer and GaP nucleation layer as part of an established GaP on Si integration process~\cite{VOLZ201137}, followed by defect-filtering layers and finished with a growth of 2.5~$\mu$m thick GaAs:Si layer, thereby establishing a virtual GaAs template with threading dislocation densities sufficiently low to preserve narrow excitonic linewidths and high single-photon purity in the telecom regime~\cite{Limame2026}. The as-grown 300~mm wafer is then cleaved into 4~cm~$\times$~4~cm dies and transferred to a second MOCVD reactor AIXTRON 200/4, optimized for high-uniformity growth of optical and electronic III-V heterostructures. The full device structure is then grown in a single continuous epitaxial sequence: a 33.5-pairs GaAs/Al$_{0.9}$Ga$_{0.1}$As DBR is grown to provide high backside reflectivity in the telecom O-band, followed by a $\lambda$-cavity incorporating an n-doped GaAs layer, an intrinsic region hosting the active emitters, and a p-doped GaAs layer, together forming a vertical p-i-n diode. Within this intrinsic region, self-assembled InGaAs QDs formed via Stranski-Krastanov growth mode and further capped with a composition-graded InGaAs strain-reducing layer, thereby redshifting the emission into the telecom O-band~\cite{Nishi1999, Paul_2015}. The QD layer is positioned at the center of the intrinsic region, placing it simultaneously at the optical antinode of the cavity for maximum light-matter interaction and equidistant from the doped contact layers for a uniform vertical electric field across the emitters, enabling deterministic control of the excitonic transitions via the QCSE. Detailed information on the epitaxial layer structure, defect-filtering layers, thicknesses, doping concentrations, and growth parameters is provided in Supplementary Information (SI) Section~1.

\subsection{Device concept and structure}\label{subsec2}

The CBG resonator consists of a central mesa housing the QD emitter, surrounded by concentric annular air trenches that form a CBG, suppressing in-plane radiation and redirecting emission into a highly directional vertical far-field pattern~\cite{Davanco_2011}. To extend this concept to electrically tunable operation, a narrow conducting ridge connects the central mesa to the outer contact, providing a conductive pathway for charge control through the reverse-biased vertical p-i-n diode Fig.~\ref{fig:1}(b). The ridge unavoidably breaks the rotational symmetry of the CBG and perturbs the in-plane refractive index profile. Its dimensions must therefore be carefully engineered to balance optical and electrical requirements: sufficiently narrow to suppress lateral waveguiding losses, yet wide enough to ensure suitably high electrical conductance despite sidewall depletion due to surface defects introduced during inductively coupled plasma-reactive ion etching (ICP-RIE). Following the design principles established by Wijitpatima \emph{et al.}~\cite{Wijitpatima2024}, we employ a ridge width of $\approx$ 150~nm, providing a practical compromise between optical integrity and electrical robustness. 

To tailor the eCBG geometry to the epitaxial layer stack and telecom O-band operation, finite-element method simulations were performed using the JCMsuite solver (JCMwave GmbH). The QD was modeled as an oscillating dipole source, and Maxwell’s equations were solved using second-order finite elements. A Bayesian optimization routine was employed to identify geometries that maximize vertically directed emission into a numerical aperture of 0.81, corresponding to the experimental collection optics. To efficiently explore the large design space, an initial reduced-dimensionality model based on a rotationally symmetric (2.5D axisymmetric) formulation was employed. This approach significantly reduces computational cost while accurately capturing the wavelength-dependent extraction characteristics of the idealized structure in the absence of symmetry-breaking elements. Using this framework, the resulting design exhibits broadband photon extraction across the telecom O-band (1260 to 1360~nm), with a PEE of $\sim$37 to 40\% and negligible Purcell enhancement, reflecting a design strategy that prioritizes spectral robustness over strong cavity confinement. Unlike high-Q microcavities such as micropillars that require tight spectral emitter-mode alignment better than 1~nm~\cite{Schlehahn2016}, the broadband eCBG architecture maintains efficient photon extraction across a large spectral range without significant reduction of source brightness. To obtain realistic device performance metrics, the full geometry of the optimized eCBG device, including the symmetry-breaking electrical ridge, was incorporated into three-dimensional simulations. These calculations account for ridge-induced perturbations of the optical mode and yield a PEE of $\sim$31\% into the collection cone at normal emission (at zero deflection). For the experimentally relevant case of a QD displaced radially by $\sim$78~nm from the eCBG-mesa center (ref. Fig.~\ref{fig:2}(b): inset), the calculated PEE decreases moderately to $\sim$26\%, indicating good tolerance to realistic emitter misalignment while maintaining efficient outcoupling~\cite{Gaur2025Scalable}. Moreover, the simulated far-field emission profile confirms highly directional outcoupling into the collection cone (see SI Fig.~S2(b)), with full structural parameters and simulation details provided in SI Section~2.

Overall, the engineered eCBG architecture retains the key optical advantages of conventional CBGs whilst integrating a vertical p-i-n diode that enables a controlled electric field to be applied across individual QD-eCBGs, providing continuous and reversible spectral tuning of the excitonic transition via the QCSE.

\subsection{Fabrication and deterministic integration}\label{subsec3}

Deterministic integration of selected QDs into eCBG resonators was achieved using a marker-based electron-beam lithography (EBL) process~\cite{Li2023}. Prior to cavity patterning, the vertical p-i-n diode structure was fabricated following an established multistep EBL protocol reported in Ref.~\cite{Wijitpatima2024}. Briefly, n-contact regions were defined by EBL and etched by ICP-RIE to expose the underlying n-type GaAs layer, followed by Ni/AuGe/Au metallization and rapid thermal annealing to form low-resistance ohmic contacts. In a Subsequent EBL step, p-contacts and alignment markers were defined and metalized using a Ti/Au deposition, after which both the metal pads were thickened with Au to ensure mechanical robustness during wire bonding. This multistep fabrication sequence provides reproducible and low-resistance electrical access while remaining fully compatible with deterministic cavity placement.

\begin{figure}[!h]
 \centering
  \includegraphics[width=1.0\textwidth]{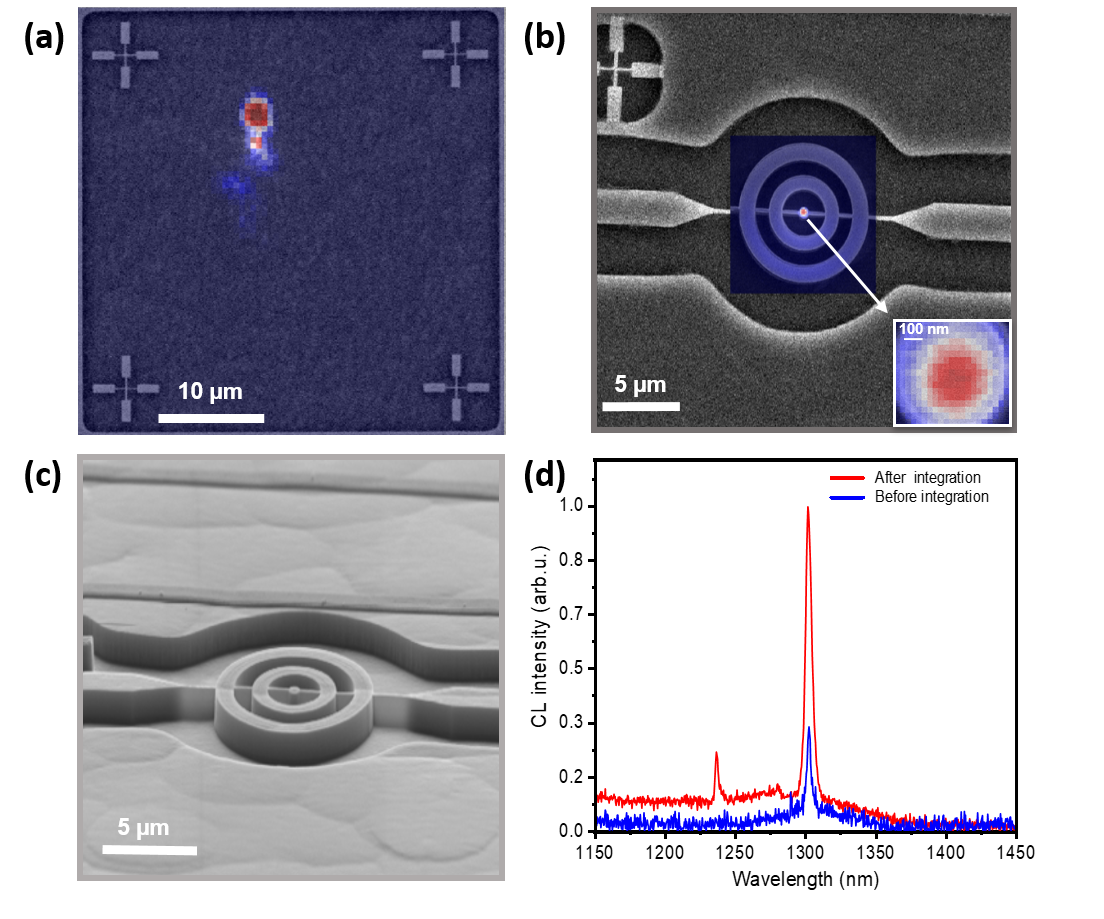}
  \caption{ \textbf{Deterministic integration of InGaAs QDs into eCBGs with enhanced O-band emission.} \textbf{(a)} CL map overlaid on a secondary-electron image, identifying spatially localized InGaAs QDs relative to alignment markers. \textbf{(b)} CL map after marker-based lithography, confirming successful integration of the selected QD into the eCBG. \textbf{(c)}  Tilted SEM image of a fabricated device, showing well-defined sidewalls and the electrical access ridge. \textbf{(d)} CL spectra at 20~K before and after integration, demonstrating enhanced emission near 1300~nm due to cavity coupling and improved vertical photon extraction.}
  \label{fig:2}
\end{figure}

Using the fabricated alignment markers, individual QDs were first localized by low-temperature cathodoluminescence (CL) mapping at 20~K under open-circuit conditions. Each marker field was scanned to generate correlated CL-SEM maps, providing precise spatial and spectral information of individual emitters as shown in Fig.~\ref{fig:2}(a). Suitable emitters were selected based on emission wavelength (in the O-band), spatial isolation, and signal contrast. The extracted spatial coordinates relative to the lithographic alignment markers were then used to precisely position the eCBG patterns defined by EBL and transferred into the device layers by ICP-RIE. Successful integration was confirmed by a post-fabrication CL map, shown in Fig.~\ref{fig:2}(b), with emission clearly localized at the cavity center, further highlighted by the inset.  A small lateral offset of 78~nm between the pre-selected QD position and the center of the fabricated cavity is attributed to temperature-induced mechanical drift of the cold finger during the initial CL mapping step~\cite{Madigawa2024}.  A representative tilted-view SEM image in Fig.~\ref{fig:2}(c) reveals smooth, near-vertical sidewalls and a well-defined electrical access ridge, confirming the high structural quality of the process. CL spectra acquired from the same QD before and after eCBG integration, shown in Fig.~\ref{fig:2}(d), show a pronounced enhancement of the emission intensity at 1310~nm due to the resonator effect of the eCBG. In addition, surface defects visible in SEM of the fabricated devices originate from threading dislocations inherent to the III-V on Si material system and are therefore practically unavoidable; however, their areal density and lateral separation are sufficiently low to leave the eCBG optical performance unaffected.

\section{Optical and quantum-optical characterization of a QD-eCBG}\label{sec2}

Following deterministic nanofabrication, the quantum devices were electrically contacted via wire bonding to a chip carrier and characterized under cryogenic conditions in a closed-cycle cryostat. Current–voltage (I-V) measurements confirm robust rectifying behavior of the fabricated p-i-n diode, exhibiting low reverse leakage and a stable operating range at both 4 K and 77 K (see SI Fig. S4), validating reliable electrical characteristics of the monolithically integrated heterostructure. Using this configuration, we performed a systematic optical and quantum-optical characterization of multiple devices at 4 K, three of which (eCBG$_1$-eCBG$_3$) are presented for detailed analysis in the following. All devices were investigated under identical experimental conditions to enable direct comparison of their optical performance and to assess the device-to-device reproducibility of the QD-eCBG platform. Device eCBG$_1$ serves as the primary reference for single-emitter performance, while eCBG$_2$ and eCBG$_3$ demonstrate electrically controlled spectral alignment of spatially separated emitters, a key requirement for multi-emitter quantum photonic architectures.

The cryogenic characterization includes $\mu$PL, bias-dependent electrical tuning, time-resolved lifetime analysis, and second-order photon autocorrelation measurements, which are presented in the following subsections. Together, these measurements provide a comprehensive evaluation of emission stability, wavelength tunability, radiative recombination dynamics, and single-photon purity under electrical control. A detailed description of the optical setup and measurement parameters is provided in SI Section 3.

\subsection{Electrical control and Stark tuning}\label{subsec1}

\begin{figure}[h]
 \centering
  \includegraphics[width=1.0\textwidth,trim=10 10 10 10,clip]{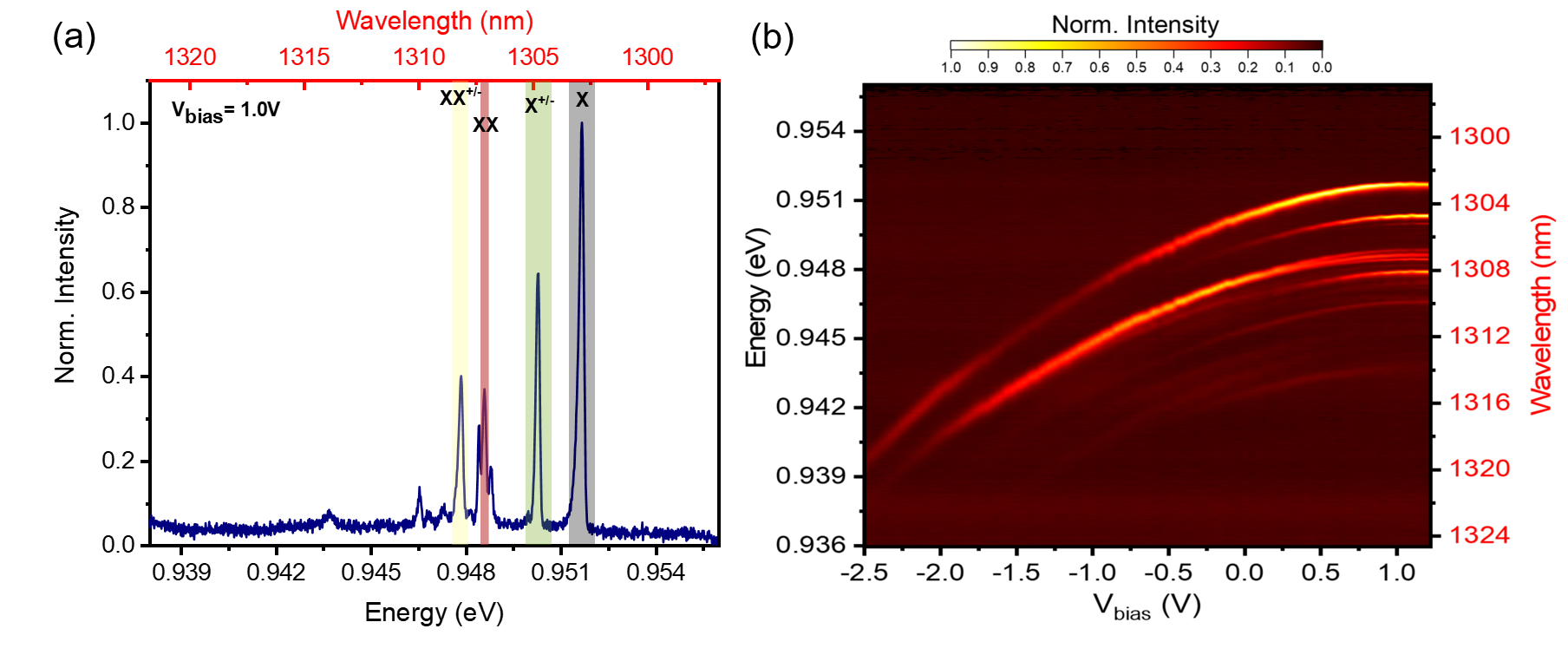}
  
  \caption{\textbf{Excitonic emission and electrically controlled wavelength tuning of a single QD in an eCBG device. (a)} $\mu$PL spectrum of eCBG$_1$ recorded at 4~K under pulsed picosecond excitation at 1225~nm, showing emission from the neutral exciton (X, black), charged exciton (X$^{\pm}$, green), biexciton (XX, red), and charged biexciton (XX$^{\pm}$, yellow). Line assignments are based on excitation-power and polarization-dependent $\mu$PL measurements. \textbf{(b)} False-color map of bias-dependent $\mu$PL spectra acquired at 4~K, demonstrating electrically controlled wavelength tuning of an excitonic transition via the QCSE.}
  
  \label{fig:3}
\end{figure}

To evaluate the optical performance of the eCBG devices under electrical control, device eCBG$_1$ was investigated under quasi-resonant pulsed excitation at 1225~nm while varying the external bias voltage ($V_{\mathrm{bias}}$) applied across the p-i-n diode. The bias-dependent $\mu$PL false-color map in Fig.~\ref{fig:3}(b) reveals systematic and continuous shifts of multiple emission lines with increasing $V_{\mathrm{bias}}$, reflecting the electrically controlled tuning of confined excitonic states via the QCSE. A representative $\mu$PL spectrum recorded at $V_{\mathrm{bias}} = 1$~V in Fig.~\ref{fig:3}(a) shows several well-resolved emission lines. 

The nature of these transitions was identified by performing power-dependent and polarization-resolved $\mu$PL measurements at the corresponding $V_{\mathrm{bias}}$, detailed information can be found in SI Section~5. Based on this analysis, the dominant lines are assigned to the neutral exciton (X, black), charged exciton (X$^{\pm}$, green), biexciton (XX, red), and charged biexciton (XX$^{\pm}$, yellow). The neutral exciton (X) emission, centered at 0.9516~eV (1302.90~nm) with a full width at half maximum of $(171 \pm 3)~\mu$eV at $V_{\mathrm{bias}}$ = 1~V, was selected for detailed analysis (Fig. 4-5). Under applied bias from 1.1~V to -2.5~V, the neutral exciton exhibits a continuous and reversible spectral tuning range of approximately 16~nm (11~meV). This large Stark shift is attributed to the deeper carrier confinement in telecom-wavelength QDs compared to their shorter-wavelength counterparts, which suppresses carrier tunneling and allows the excitonic transition to sustain significantly higher electric fields before emission quenching. Further, the InGaAs capping layer reduces the confinement of the electron–hole pair, enhancing their response to the 
applied external field and ultimately resulting in a larger cumulative energy shift. This wide tuning window provides sufficient spectral flexibility to compensate emitter-to-emitter inhomogeneity, and constitutes a record tuning range for a QD embedded in a nanophotonic structure at telecom wavelengths, surpassing previously reported values for QDs integrated into photonic devices~\cite{Wijitpatima2024, Barbiero_2025}. While tuning ranges of up to 60~nm have been demonstrated for O-band QDs in planar, unpatterned samples~\cite{Ward2014}, such values are facilitated by the absence of etched surfaces, which otherwise introduce surface-induced charge noise and limit the applicable electric field, a fundamental trade-off in nanophotonic device geometries.

The electric field across the intrinsic region was estimated assuming a uniform field distribution and is given by $F = \left(V_{\mathrm{bias}} - V_{\mathrm{bi}}\right)/t$, where $V_{\mathrm{bi}} = 1.18$~V is the built-in potential of the diode and $t = 288$~nm denotes the thickness of the intrinsic region. As  $V_{\mathrm{bias}}$ approaches $V_{\mathrm{bi}}$, the internal electric field approaches zero, corresponding to the flat-band condition of the p-i-n junction. The resulting saturation of the spectral shift indicates that the observed tuning originates from field-induced band tilting rather than carrier screening effects.

To quantify the Stark response, the excitonic transition energy was fitted using the quadratic QCSE model, $E(F) = E_0 - pF + \beta F^2$ \cite{Finley_2004, Mar_2017}, see SI, Section 5 for details. In this framework, an applied vertical electric field modifies the relative spatial alignment of the electron and hole wavefunctions along the growth axis, altering their confinement energies and Coulomb interaction and thereby shifting the optical transition energy. The field induces a relative displacement of the electron and hole wavefunctions, which results in a redshift of the excitonic emission~\cite{Barker_2000}. From this model, we extract an electron hole separation of $p/e = -(0.276\pm0.006)$~nm and a polarizability of $\beta = -(0.637\pm0.005)$~$\mu$eV/(kV/cm)$^2$ for neutral exciton (X). Details of other excitonic transitions and corresponding fits are provided in SI Section 5. The extracted values are in good agreement with previous reports on the QCSE in self-assembled InAs/GaAs QDs where typical dipole moments correspond to electron-hole separations of $\sim 0.1$-$0.5\,\text{nm}$~\cite{Fry_2000, Findeis_2001, Oulton_2002, Petruzzella_2015} and indicate an inverted electron-hole alignment, with the hole localized towards the apex and the electron towards the base of the QD, consistent with earlier experimental observations of composition-graded structures~\cite{Fry_2000}.

To quantify the PEE of the eCBG$_1$ resonator, we evaluated the detected count rates under a non-resonant excitation wavelength of 1120~nm using OPO and at a bias of 1 V. At saturation, the neutral exciton (X) exhibits an integrated superconducting nanowire single-photon detector (SNSPD) count rate of $440$~kcps, while the charged exciton contributes an additional $270$~kcps, yielding a combined count rate of $\sim 710$~kcps collected from the device. Considering the laser repetition rate of 80~MHz and the independently calibrated setup detection efficiency of $(4.1 \pm 0.3)\%$, this corresponds to a PEE of $(21.7 \pm 3.0)\%$ into the first objective. This value is consistent with the range reported for InGaAs/GaAs single-photon sources in the telecom O-band, which spans from approximately 10\% for deterministic mesa structures with backside gold mirrors~\cite{Srocka_2018}, to 11\% for QDs coupled to hybrid CBGs~\cite{Xu_2022}, and 23\% for QDs in CBG cavities~\cite{Kolatschek2021}. More recently, PEEs up to 40\% have been demonstrated for InGaAs/GaAs QDs in (non-contacted) CBG resonators on silicon substrates~\cite{Limame2026}. Importantly, the vertical p-i-n architecture provides a wide and well-defined electric-field tuning window, while the broadband eCBG cavity accommodates large spectral displacements without imposing narrow resonance constraints. Notably, the achieved PEE is obtained within an electrically contacted ridge-based device geometry, a configuration that typically introduces additional optical losses compared to passive structures, underscoring the suitability of the QD-eCBG platform as a robust and scalable architecture for electrically controllable quantum light sources.

\subsection{Time-resolved $\mu$PL studies}

\begin{figure}[h]
 \centering
  \includegraphics[width=1.05\textwidth]{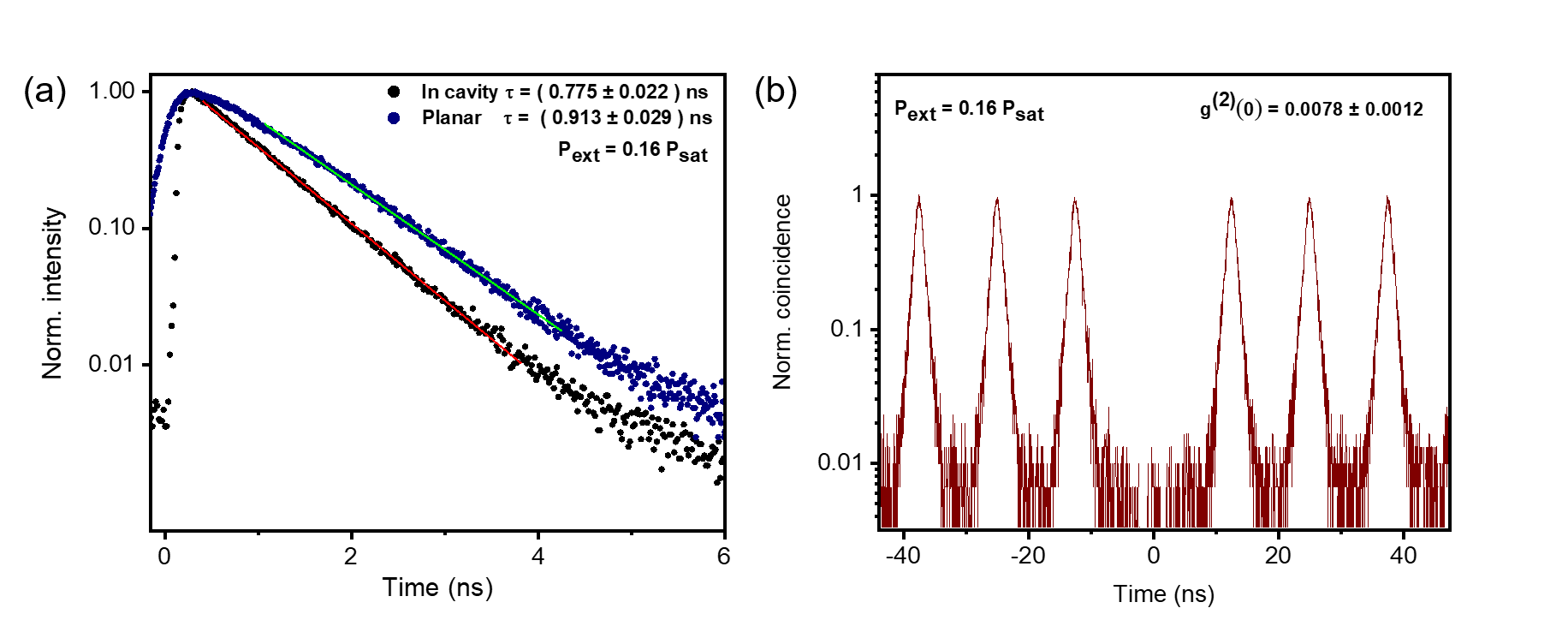}
  \caption{\textbf{Time-resolved $\mu$PL and second-order autocorrelation confirming high-purity single-photon emission from a QD-eCBG.} \textbf{(a)} Time-resolved $\mu$PL measurement showing a single-exponential decay with an extracted spontaneous emission decay time of $0.775 \pm 0.022$~ns. \textbf{(b)} Second-order autocorrelation measurement under pulsed excitation yielding fitted $g^{(2)}(0) = 0.0078 \pm 0.0012$, corresponding to a single-photon purity of $99.22 \pm 0.18\%$, comparable to state-of-the-art planar InGaAs QD single-photon sources.}
  \label{fig:4}
\end{figure}

Time-resolved $\mu$PL measurements under quasi-resonant excitation were carried out to investigate the radiative recombination dynamics of the QD embedded in device eCBG$_1$ at a bias voltage of 1 V. The measurements were conducted at a low excitation power of $P_\mathrm{ext}=0.16P_\mathrm{sat}$, well below saturation, to suppress re-excitation and carrier recapture processes and to ensure that the extracted lifetime reflects the intrinsic single-exciton radiative dynamics under applied electrical bias, independent of multi-excitonic or power-induced effects. The resulting time-resolved emission trace of the neutral exciton (X) (fig.~\ref{fig:4}(a)) is well described by a single-exponential decay, yielding a spontaneous emission decay time of $(0.775 \pm 0.022)$~ns. For comparison, nominally identical QDs in unpatterned planar regions show a lifetime of $(0.913 \pm 0.029)$~ns under identical excitation and bias conditions. The absence of a pronounced lifetime modification indicates that the radiative dynamics are largely governed by the intrinsic QD properties rather than cavity-induced effects. This is consistent with the eCBG design, which is optimized for PEE efficiency rather than Purcell enhancement. Correspondingly, finite-element simulations yield a near-unity Purcell factor (0.7 to 1.6) across the telecom O-band range, confirming that efficient photon extraction is achieved without significant modification of the spontaneous emission rate.

Bias-dependent lifetime measurements of the neutral exciton (see SI Fig.~S6(a)) further reveal a gradual increase in the radiative lifetime from  $(0.826 \pm 0.031)$~ns to  $(0.931 \pm 0.034)$~ns under a bias variation from 0.9 V to -0.9 V, which is attributed to the QCSE causing a reduction of the electron-hole wavefunction overlap through field-induced spatial separation, thereby lowering the radiative recombination rate~\cite{Mille1984}.

\subsection{Single-photon emission purity}

The single-photon purity of device eCBG$_1$ was assessed by second-order autocorrelation measurements under quasi-resonant pulsed excitation at 1225~nm using a Hanbury Brown and Twiss setup. The zero-delay autocorrelation value $g^{(2)}(0)$, which quantifies the probability of multiphoton emission within a single excitation cycle, serves as the metric for single-photon purity. Fig.~\ref{fig:4}(b)  shows the measured $g^{(2)}(\tau)$ correlation histogram acquired at a low excitation power of $P_\mathrm{ext}=0.16P_\mathrm{sat}$, where pronounced antibunching is observed at zero delay. From the fit, we extract a value of  $g^{(2)}(0)=0.0078 \pm 0.0012$, while a raw value of $g^{(2)}(0)=0.0249 \pm 0.0021$ is obtained by integrating the peak areas of the correlation histogram, corresponding to a single-photon purity exceeding $99\%$. These values are on par with the previously reported record values of $g^{(2)}(0)=0.005$ (background-corrected) for InGaAs/GaAs QDs in planar samples under p-shell excitation~\cite{Dusanowski_2017}, and $g^{(2)}(0)=0.007$ (fitted) for InGaAs/GaAs QDs integrated in CBG structures ~\cite{Limame2026}, and reflect the excellent optical performance of our O-band QD-eCBG quantum light sources.

Bias-dependent measurements (see SI Fig.~S6(a)) show consistently low $g^{(2)}(0)$ values ($<3\%$) at saturation, indicating that electrical Stark tuning does not lead to a pronounced degradation of the single-photon purity over the investigated bias range. A slight increase in $g^{(2)}(0)$ is observed at larger reverse bias, which is attributed to a reduced signal-to-background ratio at high electric fields and the corresponding reduction in emission intensity. As the applied bias approaches $V_{bi}$, the lowest $g^{(2)}(0)$ values are obtained, consistent with increased electron-hole wavefunction overlap at reduced electric fields, leading to stronger emission intensity and an improved signal-to-background ratio.

\subsection{Thermal robustness and high-temperature operation}

The thermal robustness of the QD-eCBGs was investigated through temperature-dependent optical and quantum-optical measurements on device eCBG$_1$ between 4~K and 77~K under pulsed quasi-resonant excitation at 1225~nm. All measurements were performed under saturation pump power at 4~K, 20~K, 40~K, and 77~K. This temperature range is technologically relevant, as operation near 40~K is compatible with compact closed-cycle cryocoolers~\cite{Musial_2020}, while 77~K enables liquid-nitrogen-based cooling as a cost-effective alternative to liquid-helium cryogenic systems.

\begin{figure}[h]
 \centering
  \includegraphics[width=1.0\textwidth]{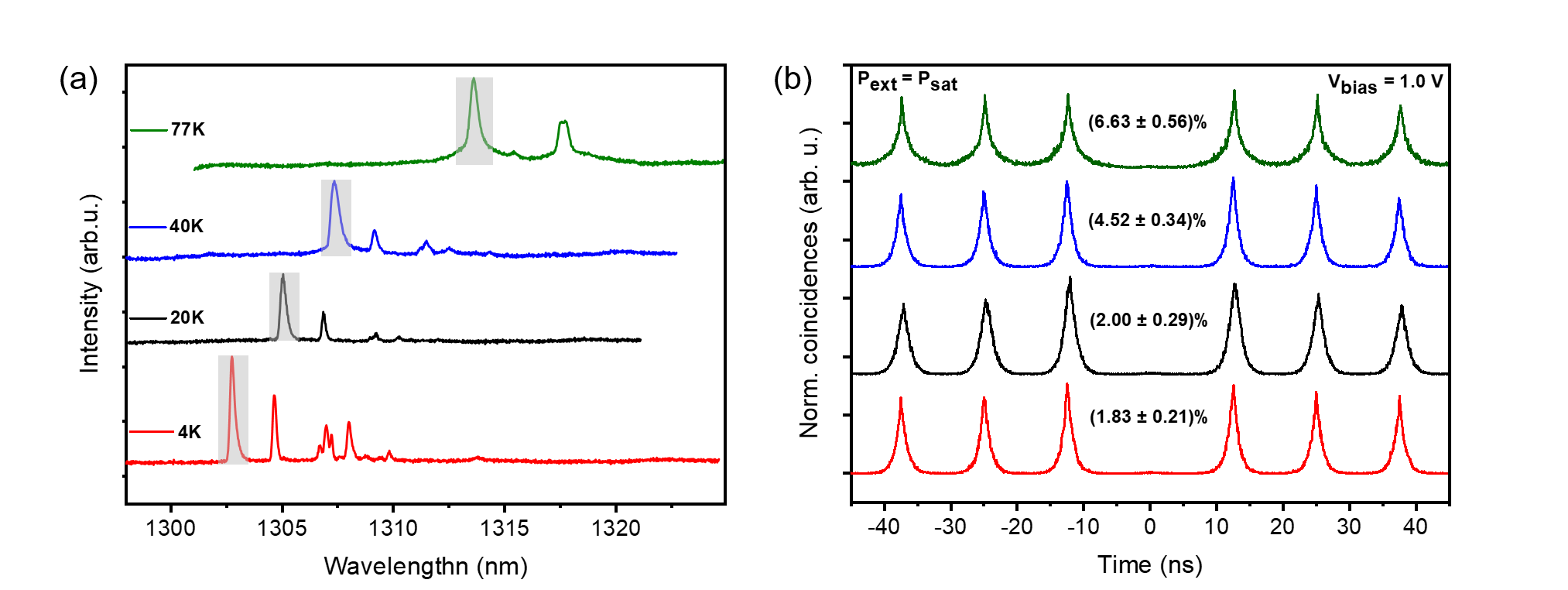}
  \caption{ \textbf{Temperature-dependent $\mu$PL and photon autocorrelation of a QD-eCBG device.} \textbf{(a)} Waterfall plot of $\mu$PL spectra recorded from an eCBG$_1$ device at temperatures of 4, 20, 40, and 77~K under pulsed excitation, showing well-resolved QD emission up to liquid-nitrogen temperature. The X transition is indicated by the gray areas. \textbf{(b)} Second-order photon autocorrelation measurements $g^{(2)}(\tau)$ of the X transition acquired at the same temperatures and at saturation pump power, demonstrating preserved single-photon emission characteristics up to 77~K.}
  
  \label{fig:5}
\end{figure}

With increasing temperature from 4~K to 77~K, the emission spectra exhibit a continuous redshift, as shown in Fig.~\ref{fig:5}(a), consistent with the temperature-dependent shrinkage of the semiconductor bandgap. Importantly, spectrally isolated single-QD emission lines remain clearly resolved up to 77~K with a reduction of emission intensity of only 28\% in the studied temperature range. This behavior indicates that carrier confinement in the QDs remains sufficiently strong to suppress thermally activated escape into the surrounding barrier or wetting layer, such that radiative recombination continues to dominate over competing non-radiative relaxation pathways~\cite{Braun_2014}. Among the observed transitions, the neutral exciton (X) remains the spectrally isolated and dominant emission feature across the full temperature range and was therefore selected for detailed quantum-optical analysis. As shown in Fig.~\ref{fig:5}(a), the emission linewidth exhibits a gradual temperature-dependent broadening. This behavior is primarily attributed to increased exciton-phonon coupling, which enhances homogeneous dephasing via phonon-induced fluctuations of the excitonic dipole phase~\cite{Bayer2002}. The absence of pronounced intensity quenching together with the nearly temperature-independent spontaneous emission (fast) decay time of about 0.9 ns (see SI, Fig. S7(b)) observed for eCBG$_1$ indicates that non-radiative recombination remains a minor channel up to 77~K.

The single-photon purity of emission was evaluated as a function of temperature using second-order photon autocorrelation measurements performed under saturation pump power, representing a stringent operating regime in which re-excitation and background contributions are significant. Pronounced antibunching is maintained across the entire investigated temperature range as shown in Fig.~\ref{fig:5}(b). The extracted raw values of $g^{(2)}(0)$ are $(1.83 \pm 0.21)\%$ at 4~K, $(2.00 \pm 0.29)\%$ at 20~K, and $(4.52 \pm 0.34)\%$ at 40~K. Even at 77~K, a raw autocorrelation value of $g^{(2)}(0) = (6.63 \pm 0.56)\%$ is obtained. The gradual increase in multi-photon probability with temperature is attributed primarily to phonon-induced pure dephasing of the excitonic zero-phonon line, which progressively broadens the emission linewidth~\cite{Thoma_2016} and increases the contribution of uncorrelated background photons within the detection window. Nevertheless, the values remain well below the classical limit across the entire temperature range, confirming true single-photon operation even at liquid-nitrogen temperature.

Taken together, the results obtained demonstrate that the strain-engineered heterostructure and vertical p-i-n architecture preserve excitonic confinement, radiative recombination, and quantum-optical purity up to 77~K. The combination of electrical tunability and stable liquid-nitrogen operation substantially relaxes cryogenic requirements, thereby significantly advancing the performance of telecom O-band QD-SPSs.

\subsection{Spectral alignment of electrically tunable QD-eCBG devices}

\begin{figure}[h]
 \centering
  \includegraphics[width=1.0\textwidth]{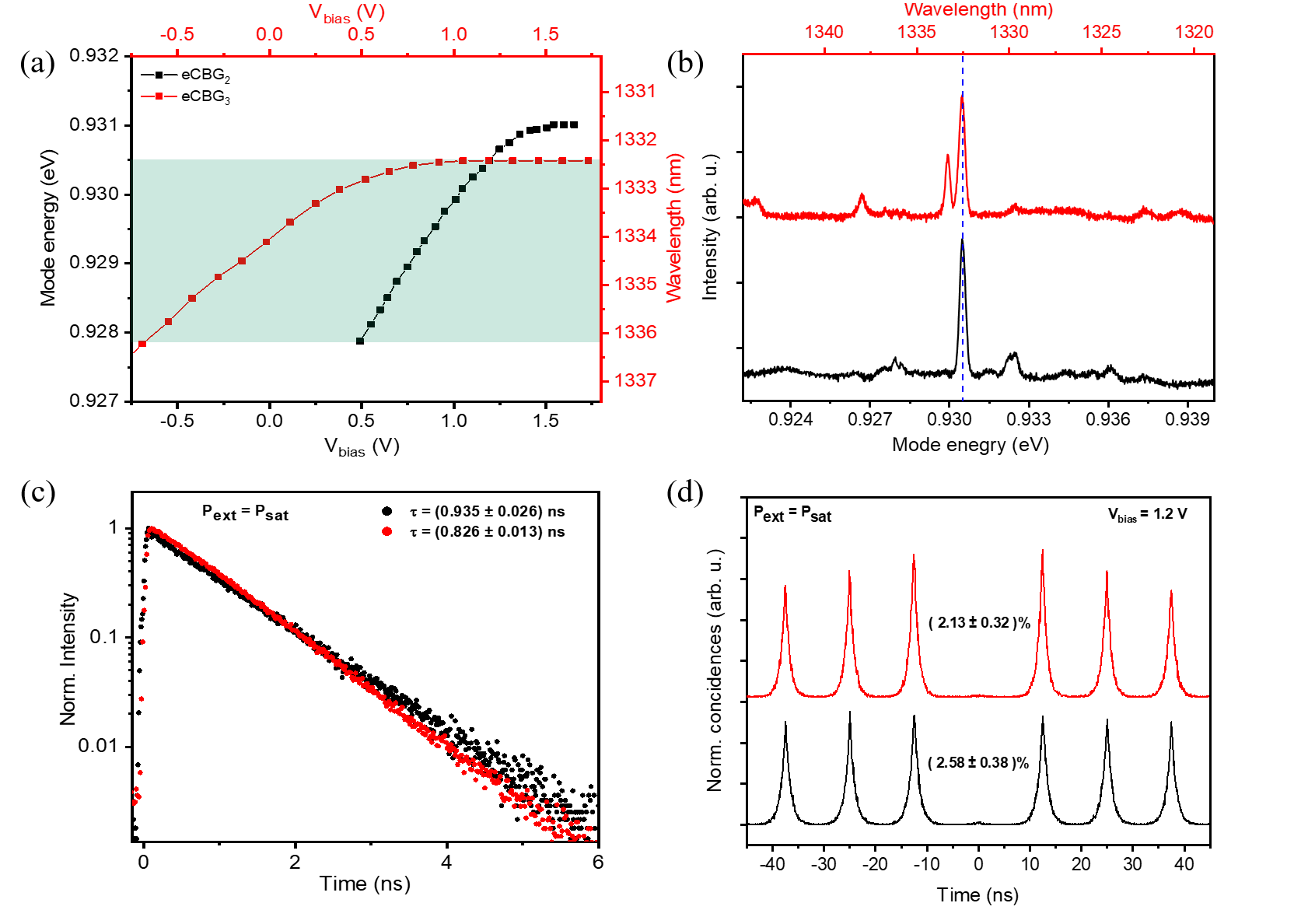}
  \caption{\textbf{Electrical tuning of two QD-eCBGs into spectral resonance with preserved single-photon emission.} \textbf{(a)} Bias-dependent tuning of two excitonic emission lines from two QD-eCBG devices, demonstrating that they can be brought into mutual resonance by electrical control. \textbf{(b)} $\mu$PL spectra recorded at the resonance bias, confirming identical emission energies. \textbf{(c)} Time-resolved $\mu$PL measurements at the resonance bias, revealing comparable radiative lifetimes of both devices. \textbf{(d)} Second-order autocorrelation measurements $g^{(2)}(\tau)$ acquired at the resonance bias, showing preserved single-photon purity for both emitters.}  
  \label{fig:6}
\end{figure}

 Achieving spectral matching between spatially separated quantum emitters is a central requirement for quantum networks based on entanglement distribution and scalable multi-emitter photonic architectures~\cite{Azuma2023}. To demonstrate a key requirement for this application, namely the fine-tuning of quantum light sources to bring them into spectral resonance, two spatially separated devices (eCBG$_2$ and eCBG$_3$) from the same array were investigated under quasi-resonant pulsed excitation. Figure~\ref{fig:6}(a) shows the bias-dependent evolution of selected excitonic transitions from both devices. At zero bias, the emitters exhibit distinct emission energies, reflecting the inherent inhomogeneity of self-assembled QDs. Upon varying the external voltage $V_\mathrm{bias}$, both transitions shift via the QCSE with different tuning slopes. Importantly, despite these distinct responses, the two emitters can be tuned into mutual resonance at a common bias of 1.2~V. The resonance condition, indicated by the dashed line, defines a shared electrical operating point at which both devices emit at identical wavelengths. The distinct Stark responses originate from dot-to-dot variations in permanent dipole moment and polarizability, reflecting differences in shape, composition, and local strain between nominally identical QDs from the same epitaxial wafer ~\cite{Fry_2000, Bennett2010}.

The corresponding $\mu$PL spectra recorded at 1.2~V bias, shown in Fig.~\ref{fig:6}(b), confirm spectral degeneracy of the two devices. To verify that electrical tuning does not compromise emitter performance, the radiative and quantum-optical properties of both emitters were evaluated at the common bias. Time-resolved $\mu$PL in Fig.~\ref{fig:6}(c) yield excitonic lifetimes of $(0.935 \pm 0.026)$~ns and $(0.826 \pm 0.013)$~ns for eCBG$_2$ and eCBG$_3$, respectively, demonstrating closely matched recombination dynamics. Such similarity in decay times is an essential requirement for interference-based protocols relying on photons emitted from independent sources. Second-order photon autocorrelation measurements acquired at the common bias Fig.~\ref{fig:6}(d) exhibit strong antibunching for both devices, with extracted raw values of $g^{(2)}(0)=(2.58 \pm 0.38)\%$ for eCBG$_2$ and $g^{(2)}(0)=(2.13 \pm 0.32)\%$ for eCBG$_3$, confirming single-photon purity under electrical tuning. This ability to achieve spectral degeneracy at a shared electrical bias, while preserving closely matched radiative dynamics and low multiphoton probability, demonstrates coordinated control of independent emitters within a common device platform. Notably, resonance is achieved through purely electrical tuning without temperature adjustment or strain engineering~\cite{Ding2016, Reindl_2016}, representing a key step toward interference-based experiments and electrically addressable multi-emitter quantum photonic circuits on silicon. Eventually, two-photon interference of remote eCBGs needs to be demonstrated for the target application, but this is beyond the scope of the present work. 

\section{Conclusion} 

This work establishes QD-eCBG devices as a silicon-compatible platform for high-purity single-photon generation at telecommunication wavelengths. By combining large Stark tuning, bias-stable quantum-optical performance, and operation up to liquid-nitrogen temperature within a unified architecture, the platform addresses several longstanding constraints of telecom quantum-dot sources like spectral inhomogeneity, tuning-induced performance degradation, and the requirement for deep cryogenic operation.

A central advance of this work is the realization of large and reversible electrical Stark tuning in the telecom O-band without compromising radiative dynamics or photon statistics. The achieved tuning window of~16 nm (11 meV) is sufficient to compensate intrinsic emitter-to-emitter spectral variations while maintaining high single-photon purity, with $g^{(2)}(0)$ values ($<3\%$) at saturation across the full accessible bias range. In contrast to global approaches such as temperature~\cite{Reindl_2016} or strain tuning~\cite{Ding2016}, this electrical bias approach provides local, dynamically addressable control of individual emitters complemented by an efficient photon extraction of $\approx$21.7\% in a ridge-based architecture, thereby enabling a scalable route toward multi-emitter integration within complex photonic circuits. Operation up to liquid-nitrogen temperature further enhances the practical relevance of the platform. The preservation of stable recombination dynamics and high single-photon purity at 77 K with $g^{(2)}(0)$ value of 6.6\% indicates that elevated-temperature operation can be achieved without introducing significant charge noise or thermally activated background processes, representing a substantial reduction in system complexity compared to liquid-helium-based platforms.

Scalability is further demonstrated through electrical spectral alignment of two spatially separated QD-eCBGs within the same device array. Bringing independent QDs into mutual resonance at a shared electrical operating point, while preserving comparable radiative lifetimes and high single-photon purity, with $g^{(2)}(0)$ values ($<2.6\%$) under saturation for both emitters, fulfills a key prerequisite for interference-based quantum photonic protocols. The broadband eCBG design supports this coordinated control by providing efficient photon extraction with tolerance to spectral shifts, thereby enabling simultaneous tunability and stability within a unified device architecture. Collectively, these results advance electrically tunable telecom QD single-photon sources from proof-of-principle demonstrations toward scalable and technologically deployable quantum photonic systems.

\section*{Supplementary information}

The Supplementary Information provides comprehensive details of the epitaxial layer structure, eCBG resonator design parameters, experimental setup, and current-voltage characteristics of the fabricated diode, along with bias-dependent optical and quantum-optical measurements and temperature-dependent optical characterization of eCBG$_1$.

\section*{Acknowledgements}

The authors thank Aris Koulas-Simos for their valuable input in scientific discussions and Setthanat Wijitpatima for helpful discussions regarding numeric modelling and device fabrication. The authors would also like to thank Kathrin Schatke, Praphat Sonka, Heike Oppermann, and Stefan Bock for their expert technical support.

\section*{Declarations}
\begin{itemize}
\item Funding:
This work was funded by the German Research Foundation via the projects Re2974/23-1, Re2974/33-1 and INST 131/795-1 320 FUGG, from the Federal Ministry of Research, Technology and Space (BMFTR) via the project MultiCoreSPS (Grant No. 16KIS1819K), and from Berlin Quantum (BQ).

\item Conflict of interest:
The authors declare no conflicts of interest.

\item Data availability:
The data that support the findings of this study are available from the corresponding authors upon reasonable request.

\item Author contribution: S.T. led the sample growth, device fabrication, manuscript preparation, data acquisition, and analysis, and contributed equally to the conceptualization, experimental investigation, and methodology development. K.G. contributed equally to the conceptualization, methodology development, experimental investigation, and manuscript preparation. He also led the simulations and supported sample growth, device fabrication, and analysis. P.M. did the wire bonding and supported methodology development, device fabrication, and numerical simulations. I.L. supported methodology development, sample growth, and experimental investigation. A.K. supported the experimental investigation. P.L. and K.V. provided the III-V grown silicon templates. S.R. led the conceptualization, funding acquisition, project administration, and supervision. He also co-authored the original draft and contributed equally to the review and editing of the manuscript.
\end{itemize}


\begin{thebibliography}{10}
\providecommand{\url}[1]{\texttt{#1}}
\providecommand{\urlprefix}{URL }

\bibitem{Su2020}
X.~Su, M.~Wang, Z.~Yan, X.~Jia, C.~Xie, K.~Peng,
\newblock \emph{Science China Information Sciences} \textbf{2020}, \emph{63}, 8
  180503.

\bibitem{Muller2014}
M.~M{\"u}ller, S.~Bounouar, K.~D. J{\"o}ns, M.~Gl{\"a}ssl, P.~Michler,
\newblock \emph{Nature Photonics} \textbf{2014}, \emph{8}, 3 224.

\bibitem{Ding2016}
X.~Ding, Y.~He, Z.-C. Duan, N.~Gregersen, M.-C. Chen, S.~Unsleber, S.~Maier,
  C.~Schneider, M.~Kamp, S.~H\"ofling, C.-Y. Lu, J.-W. Pan,
\newblock \emph{Phys. Rev. Lett.} \textbf{2016}, \emph{116} 020401.

\bibitem{Huber2017}
D.~Huber, M.~Reindl, Y.~Huo, H.~Huang, J.~S. Wildmann, O.~G. Schmidt,
  A.~Rastelli, R.~Trotta,
\newblock \emph{Nature Communications} \textbf{2017}, \emph{8}, 1 15506.

\bibitem{Senellart2017}
P.~Senellart, G.~Solomon, A.~White,
\newblock \emph{Nature Nanotechnology} \textbf{2017}, \emph{12}, 11 1026.

\bibitem{Liu2019}
J.~Liu, R.~Su, Y.~Wei, B.~Yao, S.~F. C.~d. Silva, Y.~Yu, J.~Iles-Smith,
  K.~Srinivasan, A.~Rastelli, J.~Li, X.~Wang,
\newblock \emph{Nature Nanotechnology} \textbf{2019}, \emph{14}, 6 586.

\bibitem{Uppu_2020_SA}
R.~Uppu, F.~T. Pedersen, Y.~Wang, C.~T. Olesen, C.~Papon, X.~Zhou, L.~Midolo,
  S.~Scholz, A.~D. Wieck, A.~Ludwig, P.~Lodahl,
\newblock \emph{Science Advances} \textbf{2020}, \emph{6}, 50 eabc8268.

\bibitem{Heindel2023}
T.~Heindel, J.-H. Kim, N.~Gregersen, A.~Rastelli, S.~Reitzenstein,
\newblock \emph{Adv. Opt. Photon.} \textbf{2023}, \emph{15}, 3 613.

\bibitem{Wang2025}
H.~Wang, T.~C. Ralph, J.~J. Renema, C.-Y. Lu, J.-W. Pan,
\newblock \emph{Nature Materials} \textbf{2025}.

\bibitem{Somaschi2016}
N.~Somaschi, V.~Giesz, L.~De~Santis, J.~C. Loredo, M.~P. Almeida, G.~Hornecker,
  S.~L. Portalupi, T.~Grange, C.~Ant{\'o}n, J.~Demory, C.~G{\'o}mez, I.~Sagnes,
  N.~D. Lanzillotti-Kimura, A.~Lema{\'i}tre, A.~Auffeves, A.~G. White,
  L.~Lanco, P.~Senellart,
\newblock \emph{Nature Photonics} \textbf{2016}, \emph{10}, 5 340.

\bibitem{Wang2019}
H.~Wang, H.~Hu, T.-H. Chung, J.~Qin, X.~Yang, J.-P. Li, R.-Z. Liu, H.-S. Zhong,
  Y.-M. He, X.~Ding, Y.-H. Deng, Q.~Dai, Y.-H. Huo, S.~H\"ofling, C.-Y. Lu,
  J.-W. Pan,
\newblock \emph{Phys. Rev. Lett.} \textbf{2019}, \emph{122} 113602.

\bibitem{Tomm2021}
N.~Tomm, A.~Javadi, N.~O. Antoniadis, D.~Najer, M.~C. L{\"o}bl, A.~R. Korsch,
  R.~Schott, S.~R. Valentin, A.~D. Wieck, A.~Ludwig, R.~J. Warburton,
\newblock \emph{Nature Nanotechnology} \textbf{2021}, \emph{16}, 4 399.

\bibitem{Lee2021}
C.-M. Lee, M.~A. Buyukkaya, S.~Harper, S.~Aghaeimeibodi, C.~J.~K. Richardson,
  E.~Waks,
\newblock \emph{Nano Letters} \textbf{2021}, \emph{21}, 1 323.

\bibitem{Kolatschek2021}
S.~Kolatschek, C.~Nawrath, S.~Bauer, J.~Huang, J.~Fischer, R.~Sittig,
  M.~Jetter, S.~L. Portalupi, P.~Michler,
\newblock \emph{Nano Letters} \textbf{2021}, \emph{21}, 18 7740.

\bibitem{Holewa2024}
P.~Holewa, D.~A. Vajner, E.~Zieba-Ostoj, M.~Wasiluk, B.~Gaal,
  A.~Sakanas, M.~G. Mikulicz, P.~Mrowinski, B.~Krajnik, M.~Xiong,
  K.~Yvind, N.~Gregersen, A.~Musial, A.~Huck, T.~Heindel, M.~Syperek,
  E.~Semenova,
\newblock \emph{Nature Communications} \textbf{2024}, \emph{15}, 1 3358.

\bibitem{Hauser2026}
N.~Hauser, M.~Bayerbach, J.~Kaupp, Y.~Reum, G.~Peniakov, J.~Michl, M.~Kamp,
  T.~Huber-Loyola, A.~T. Pfenning, S.~H{\"o}fling, S.~Barz,
\newblock \emph{Nature Communications} \textbf{2026}, \emph{17}, 1 537.

\bibitem{Gisin2002}
N.~Gisin, G.~Ribordy, W.~Tittel, H.~Zbinden,
\newblock \emph{Rev. Mod. Phys.} \textbf{2002}, \emph{74} 145.

\bibitem{Agarwal2021}
\emph{Optical Fibers}, chapter~2, 21--65,
\newblock John Wiley \& Sons, Ltd,
\newblock ISBN 9781119737391, \textbf{2021}.

\bibitem{ZaskePRL2012}
S.~Zaske, A.~Lenhard, C.~A. Ke\ss{}ler, J.~Kettler, C.~Hepp, C.~Arend,
  R.~Albrecht, W.-M. Schulz, M.~Jetter, P.~Michler, C.~Becher,
\newblock \emph{Phys. Rev. Lett.} \textbf{2012}, \emph{109} 147404.

\bibitem{Singh2019}
A.~Singh, Q.~Li, S.~Liu, Y.~Yu, X.~Lu, C.~Schneider, S.~H\"{o}fling, J.~Lawall,
  V.~Verma, R.~Mirin, S.~W. Nam, J.~Liu, K.~Srinivasan,
\newblock \emph{Optica} \textbf{2019}, \emph{6}, 5 563.

\bibitem{Nishi1999}
K.~Nishi, H.~Saito, S.~Sugou, J.-S. Lee,
\newblock \emph{Applied Physics Letters} \textbf{1999}, \emph{74}, 8 1111.

\bibitem{HOSPODKOVA2007}
A.~Hospodková, E.~Hulicius, J.~Oswald, J.~Pangrác, T.~Mates, K.~Kuldová,
  K.~Melichar, T.~Šimeček,
\newblock \emph{Journal of Crystal Growth} \textbf{2007}, \emph{298} 582.

\bibitem{Paul_2015}
M.~Paul, J.~Kettler, K.~Zeuner, C.~Clausen, M.~Jetter, P.~Michler,
\newblock \emph{Applied Physics Letters} \textbf{2015}, \emph{106}, 12 122105.

\bibitem{Benyoucef_2013}
M.~Benyoucef, M.~Yacob, J.~P. Reithmaier, J.~Kettler, P.~Michler,
\newblock \emph{Applied Physics Letters} \textbf{2013}, \emph{103}, 16 162101.

\bibitem{Miyazawa_2016}
T.~Miyazawa, K.~Takemoto, Y.~Nambu, S.~Miki, T.~Yamashita, H.~Terai,
  M.~Fujiwara, M.~Sasaki, Y.~Sakuma, M.~Takatsu, T.~Yamamoto, Y.~Arakawa,
\newblock \emph{Applied Physics Letters} \textbf{2016}, \emph{109}, 13 132106.

\bibitem{Musial_2020}
A.~Musiał, K.~Zolnacz, N.~Srocka, O.~Kravets, J.~Große, J.~Olszewski,
  K.~Poturaj, G.~Wojcik, P.~Mergo, K.~Dybka, M.~Dyrkacz, M.~Dłubek,
  K.~Lauritsen, A.~Bülter, P.-I. Schneider, L.~Zschiedrich, S.~Burger,
  S.~Rodt, W.~Urbanczyk, G.~Sek, S.~Reitzenstein,
\newblock \emph{Advanced Quantum Technologies} \textbf{2020}, \emph{3}, 6
  2000018.

\bibitem{Gao_2022}
T.~Gao, L.~Rickert, F.~Urban, J.~Große, N.~Srocka, S.~Rodt, A.~Musial,
  K.~Zolnacz, P.~Mergo, K.~Dybka, W.~Urbanczyk, G.~Sek, S.~Burger,
  S.~Reitzenstein, T.~Heindel,
\newblock \emph{Applied Physics Reviews} \textbf{2022}, \emph{9}, 1 011412.

\bibitem{Yang2024}
J.~Yang, Z.~Rao, C.~Song, M.~Rao, Z.~Zheng, L.~Liu, X.~Peng, Y.~Yu, S.~Yu,
\newblock \emph{Photonics Research} \textbf{2024}, \emph{12}, 10 2130.

\bibitem{Katsumi2019}
R.~Katsumi, Y.~Ota, A.~Osada, T.~Yamaguchi, T.~Tajiri, M.~Kakuda, S.~Iwamoto,
  H.~Akiyama, Y.~Arakawa,
\newblock \emph{Applied Physics Letters Photonics} \textbf{2019}, \emph{4}, 3
  036105.

\bibitem{Vijayan2024}
P.~Vijayan, R.~Joos, M.~Werner, J.~Hirlinger-Alexander, M.~Seibold, S.~Vollmer,
  R.~Sittig, S.~Bauer, F.~Braun, S.~L. Portalupi, M.~Jetter, P.~Michler,
\newblock \emph{Materials for Quantum Technology} \textbf{2024}, \emph{4}, 1
  016301.

\bibitem{Luxmoore2013}
I.~J. Luxmoore, R.~Toro, O.~D. Pozo-Zamudio, N.~A. Wasley, E.~A. Chekhovich,
  A.~M. Sanchez, R.~Beanland, A.~M. Fox, M.~S. Skolnick, H.~Y. Liu, A.~I.
  Tartakovskii,
\newblock \emph{Scientific Reports} \textbf{2013}, \emph{3}, 1 1239.

\bibitem{Tang2014}
M.~Tang, S.~Chen, J.~Wu, Q.~Jiang, V.~G. Dorogan, M.~Benamara, Y.~I. Mazur,
  G.~J. Salamo, A.~Seeds, H.~Liu,
\newblock \emph{Opt. Express} \textbf{2014}, \emph{22}, 10 11528.

\bibitem{Limame2024}
I.~Limame, P.~Ludewig, C.-W. Shih, M.~Hohn, C.~C. Palekar, W.~Stolz,
  S.~Reitzenstein,
\newblock \emph{Optica Quantum} \textbf{2024}, \emph{2}, 2 117.

\bibitem{Paesani2020}
S.~Paesani, M.~Borghi, S.~Signorini, A.~Ma{\"i}nos, L.~Pavesi, A.~Laing,
\newblock \emph{Nature Communications} \textbf{2020}, \emph{11}, 1 2505.

\bibitem{Olbrich_2017}
F.~Olbrich, J.~Kettler, M.~Bayerbach, M.~Paul, J.~Höschele, S.~L. Portalupi,
  M.~Jetter, P.~Michler,
\newblock \emph{Journal of Applied Physics} \textbf{2017}, \emph{121}, 18
  184302.

\bibitem{Holewa2020TelecomQD}
P.~Holewa, M.~G. Mikulicz, A.~Musial, N.~Srocka, D.~Quandt, A.~Strittmatter,
  S.~Rodt, S.~Reitzenstien, G.~Sek,
\newblock \emph{Scientific Reports} \textbf{2020}, \emph{10} 21816.

\bibitem{Schlehahn_2015}
A.~Schlehahn, L.~Krüger, M.~Gschrey, J.-H. Schulze, S.~Rodt, A.~Strittmatter,
  T.~Heindel, S.~Reitzenstein,
\newblock \emph{Review of Scientific Instruments} \textbf{2015}, \emph{86}, 1
  013113.

\bibitem{Davanco_2011}
M.~Davanço, M.~T. Rakher, D.~Schuh, A.~Badolato, K.~Srinivasan,
\newblock \emph{Applied Physics Letters} \textbf{2011}, \emph{99}, 4 041102.

\bibitem{Sapienza2015}
L.~Sapienza, M.~Davan\c{c}o, A.~Badolato, K.~Srinivasan,
\newblock \emph{Nature Communications} \textbf{2015}, \emph{6}, 1.

\bibitem{Gaur2025Scalable}
K.~Gaur, A.~Barua, S.~Tripathi, L.~J. Roche, S.~Wilksen, A.~Steinhoff,
  S.~Baraz, N.~Nitin, C.~C. Palekar, A.~Koulas-Simos, I.~Limame, P.~Mudi,
  S.~Rodt, C.~Gies, S.~Reitzenstein,
\newblock \emph{Light: Science {\&} Applications} \textbf{2026}, \emph{15}, 1
  260.

\bibitem{Baier_2004}
M.~H. Baier, S.~Watanabe, E.~Pelucchi, E.~Kapon,
\newblock \emph{Applied Physics Letters} \textbf{2004}, \emph{84}, 11 1943.

\bibitem{Schneider_2008}
C.~Schneider, M.~Strauß, T.~Sünner, A.~Huggenberger, D.~Wiener,
  S.~Reitzenstein, M.~Kamp, S.~Höfling, A.~Forchel,
\newblock \emph{Applied Physics Letters} \textbf{2008}, \emph{92}, 18 183101.

\bibitem{Gaur2025MQT}
K.~Gaur, P.~Mudi, P.~Klenovsky, S.~Reitzenstein,
\newblock \emph{Materials for Quantum Technology} \textbf{2025}, \emph{5}, 2
  022002.

\bibitem{Limame2026SCQDs}
I.~Limame, C.-W. Shih, K.~Gaur, M.~Podhorsk{\`y}, S.~Tripathi, S.~Wijitpatima,
  A.~Koulas-Simos, C.~C. Palekar, P.~Klenovsk{\`y}, S.~Reitzenstein,
\newblock \emph{arXiv preprint arXiv:2603.23392} \textbf{2026}.

\bibitem{Mille1984}
D.~A.~B. Miller, D.~S. Chemla, T.~C. Damen, A.~C. Gossard, W.~Wiegmann, T.~H.
  Wood, C.~A. Burrus,
\newblock \emph{Phys. Rev. Lett.} \textbf{1984}, \emph{53} 2173.

\bibitem{Finley_2004}
J.~J. Finley, M.~Sabathil, P.~Vogl, G.~Abstreiter, R.~Oulton, A.~I.
  Tartakovskii, D.~J. Mowbray, M.~S. Skolnick, S.~L. Liew, A.~G. Cullis,
  M.~Hopkinson,
\newblock \emph{Phys. Rev. B} \textbf{2004}, \emph{70} 201308.

\bibitem{Bennett2010}
A.~J. Bennett, R.~B. Patel, J.~Skiba-Szymanska, C.~A. Nicoll, I.~Farrer, D.~A.
  Ritchie, A.~J. Shields,
\newblock \emph{Applied Physics Letters} \textbf{2010}, \emph{97}, 3 031104.

\bibitem{Wijitpatima2024}
S.~Wijitpatima, N.~Auler, P.~Mudi, T.~Funk, A.~Barua, B.~Shrestha, J.~Schall,
  I.~Limame, S.~Rodt, D.~Reuter, S.~Reitzenstein,
\newblock \emph{ACS Nano} \textbf{2024}, \emph{18}, 46 31834.

\bibitem{Azuma2023}
K.~Azuma, S.~E. Economou, D.~Elkouss, P.~Hilaire, L.~Jiang, H.-K. Lo,
  I.~Tzitrin,
\newblock \emph{Rev. Mod. Phys.} \textbf{2023}, \emph{95} 045006.

\bibitem{VOLZ201137}
K.~Volz, A.~Beyer, W.~Witte, J.~Ohlmann, I.~Németh, B.~Kunert, W.~Stolz,
\newblock \emph{Journal of Crystal Growth} \textbf{2011}, \emph{315}, 1 37,
  15th International Conference on Metalorganic Vapor Phase Epitaxy
  (ICMOVPE-XV).

\bibitem{Limame2026}
I.~Limame, P.~Ludewig, A.~Koulas-Simos, C.~C. Palekar, J.~Donges, C.-W. Shih,
  K.~Gaur, S.~Tripathi, S.~Rodt, W.~Stolz, K.~Volz, S.~Reitzenstein,
\newblock \emph{APL Photonics} \textbf{2026}, \emph{11}, 4.

\bibitem{Schlehahn2016}
A.~Schlehahn, A.~Thoma, P.~Munnelly, M.~Kamp, S.~Höfling, T.~Heindel,
  C.~Schneider, S.~Reitzenstein,
\newblock \emph{APL Photonics} \textbf{2016}, \emph{1}, 1 011301.

\bibitem{Li2023}
S.~Li, Y.~Yang, J.~Schall, M.~von Helversen, C.~Palekar, H.~Liu, L.~Roche,
  S.~Rodt, H.~Ni, Y.~Zhang, Z.~Niu, S.~Reitzenstein,
\newblock \emph{ACS Photonics} \textbf{2023}, \emph{10}, 8 2846.

\bibitem{Madigawa2024}
A.~A. Madigawa, J.~N. Donges, B.~Ga{\'a}l, S.~Li, M.~A. Jacobsen, H.~Liu,
  D.~Dai, X.~Su, X.~Shang, H.~Ni, J.~Schall, S.~Rodt, Z.~Niu, N.~Gregersen,
  S.~Reitzenstein, B.~Munkhbat,
\newblock \emph{ACS Photonics} \textbf{2024}, \emph{11}, 3 1012.

\bibitem{Barbiero_2025}
A.~Barbiero, G.~Shooter, J.~Skiba-Szymanska, J.~Huang, L.~Ravi, J.~I. Davies,
  B.~Ramsay, D.~J.~P. Ellis, A.~J. Shields, T.~M{\"u}ller, R.~M. Stevenson,
\newblock \emph{ACS Photonics} \textbf{2025}, \emph{12}, 12 6607.

\bibitem{Ward2014}
M.~B. Ward, M.~C. Dean, R.~M. Stevenson, A.~J. Bennett, D.~J.~P. Ellis,
  K.~Cooper, I.~Farrer, C.~A. Nicoll, D.~A. Ritchie, A.~J. Shields,
\newblock \emph{Nature Communications} \textbf{2014}, \emph{5}, 1 3316.

\bibitem{Mar_2017}
J.~D. Mar, J.~J. Baumberg, X.~L. Xu, A.~C. Irvine, D.~A. Williams,
\newblock \emph{Phys. Rev. B} \textbf{2017}, \emph{95} 201304.

\bibitem{Barker_2000}
J.~A. Barker, E.~P. O'Reilly,
\newblock \emph{Phys. Rev. B} \textbf{2000}, \emph{61} 13840.

\bibitem{Fry_2000}
P.~W. Fry, I.~E. Itskevich, D.~J. Mowbray, M.~S. Skolnick, J.~J. Finley, J.~A.
  Barker, E.~P. O'Reilly, L.~R. Wilson, I.~A. Larkin, P.~A. Maksym,
  M.~Hopkinson, M.~Al-Khafaji, J.~P.~R. David, A.~G. Cullis, G.~Hill, J.~C.
  Clark,
\newblock \emph{Phys. Rev. Lett.} \textbf{2000}, \emph{84} 733.

\bibitem{Findeis_2001}
F.~Findeis, M.~Baier, E.~Beham, A.~Zrenner, G.~Abstreiter,
\newblock \emph{Applied Physics Letters} \textbf{2001}, \emph{78}, 19 2958.

\bibitem{Oulton_2002}
R.~Oulton, J.~J. Finley, A.~D. Ashmore, I.~S. Gregory, D.~J. Mowbray, M.~S.
  Skolnick, M.~J. Steer, S.-L. Liew, M.~A. Migliorato, A.~J. Cullis,
\newblock \emph{Phys. Rev. B} \textbf{2002}, \emph{66} 045313.

\bibitem{Petruzzella_2015}
M.~Petruzzella, T.~Xia, F.~Pagliano, S.~Birindelli, L.~Midolo, Z.~Zobenica,
  L.~H. Li, E.~H. Linfield, A.~Fiore,
\newblock \emph{Applied Physics Letters} \textbf{2015}, \emph{107}, 14 141109.

\bibitem{Srocka_2018}
N.~Srocka, A.~Musial, P.-I. Schneider, P.~Mrowinski, P.~Holewa, S.~Burger,
  D.~Quandt, A.~Strittmatter, S.~Rodt, S.~Reitzenstein, G.~Sek,
\newblock \emph{AIP Advances} \textbf{2018}, \emph{8}, 8 085205.

\bibitem{Xu_2022}
S.-W. Xu, Y.-M. Wei, R.-B. Su, X.-S. Li, P.-N. Huang, S.-F. Liu, X.-Y. Huang,
  Y.~Yu, J.~Liu, X.-H. Wang,
\newblock \emph{Photon. Res.} \textbf{2022}, \emph{10}, 8 B1.

\bibitem{Dusanowski_2017}
{\L}.~Dusanowski, P.~Holewa, A.~Marynski, A.~Musial, T.~Heuser,
  N.~Srocka, D.~Quandt, A.~Strittmatter, S.~Rodt, J.~Misiewicz,
  S.~Reitzenstein, G.~Sek,
\newblock \emph{Opt. Express} \textbf{2017}, \emph{25}, 25 31122.

\bibitem{Braun_2014}
T.~Braun, C.~Schneider, S.~Maier, R.~Igusa, S.~Iwamoto, A.~Forchel,
  S.~Höfling, Y.~Arakawa, M.~Kamp,
\newblock \emph{AIP Advances} \textbf{2014}, \emph{4}, 9 097128.

\bibitem{Bayer2002}
M.~Bayer, A.~Forchel,
\newblock \emph{Phys. Rev. B} \textbf{2002}, \emph{65} 041308.

\bibitem{Thoma_2016}
A.~Thoma, P.~Schnauber, M.~Gschrey, M.~Seifried, J.~Wolters, J.-H. Schulze,
  A.~Strittmatter, S.~Rodt, A.~Carmele, A.~Knorr, T.~Heindel, S.~Reitzenstein,
\newblock \emph{Phys. Rev. Lett.} \textbf{2016}, \emph{116} 033601.

\bibitem{Reindl_2016}
M.~Reindl, K.~D. J{\"o}ns, D.~Huber, C.~Schimpf, Y.~Huo, V.~Zwiller,
  A.~Rastelli, R.~Trotta,
\newblock \emph{Nano Letters} \textbf{2017}, \emph{17}, 7 4090.

\end{thebibliography}

\end{document}


\pagestyle{fancy}
\rhead{\includegraphics[width=2.5cm]{vch-logo.png}}

\title{Stark-tunable O‑band single-photon sources based on deterministically fabricated quantum dot - circular Bragg gratings on silicon: Supplementary Information}
\maketitle

\vspace*{2mm}

\author{Sarthak Tripathi$^{1}$},
\author{Kartik Gaur$^{1}$},
\author{Priyabrata Mudi$^{1}$},
\author{Peter Ludewig$^{2,3}$},
\author{Alexander Kosarev$^{1}$},
\author{Kerstin Volz$^{2,3}$},
\author{Imad Limame$^{1}$},
\author{and Stephan Reitzenstein$^{1}$*}

\vspace*{2mm}

\begin{affiliations}

{$^1$Institut für Physik und Astronomie, Technische Universit\"at Berlin, Hardenbergstrasse 36, 10623 Berlin, Germany}

{$^2$mar.quest - Marburg Center for Quantum Materials and Sustainable Technologies, Philipps-Universität Marburg, Hans Meerwein Str. 6, 35032 Marburg, Germany}

{$^3$Department of Physics, Philipps-Universität Marburg, Structure and Technology Research Lab, Hans Meerwein Str.~6, 35032 Marburg, Germany}

\end{affiliations}

*Corresponding author: stephan.reitzenstein@physik.tu-berlin.de

\keywords{quantum dot, Stark shift, circular Bragg grating, single-photon source, telecom O-band, III-V on silicon}

\vspace{0.3cm}

This Supplementary Information (SI) provides detailed support to the main manuscript, including a comprehensive description of the epitaxial layer structure and electrically contacted circular Bragg grating (eCBG) resonator design, together with numerical simulation parameters. It further outlines the experimental setup, as well as the electrical characteristics of the fabricated p-i-n diode. In addition, extended optical and quantum-optical measurements are presented, including bias-dependent studies and temperature-dependent characterization of eCBG$_{1}$, providing deeper insight into device performance and stability.

\section{Epitaxial growth and layer design}\label{sec1}

\begin{figure}[h]
 \centering
  \includegraphics[width=0.95\textwidth]{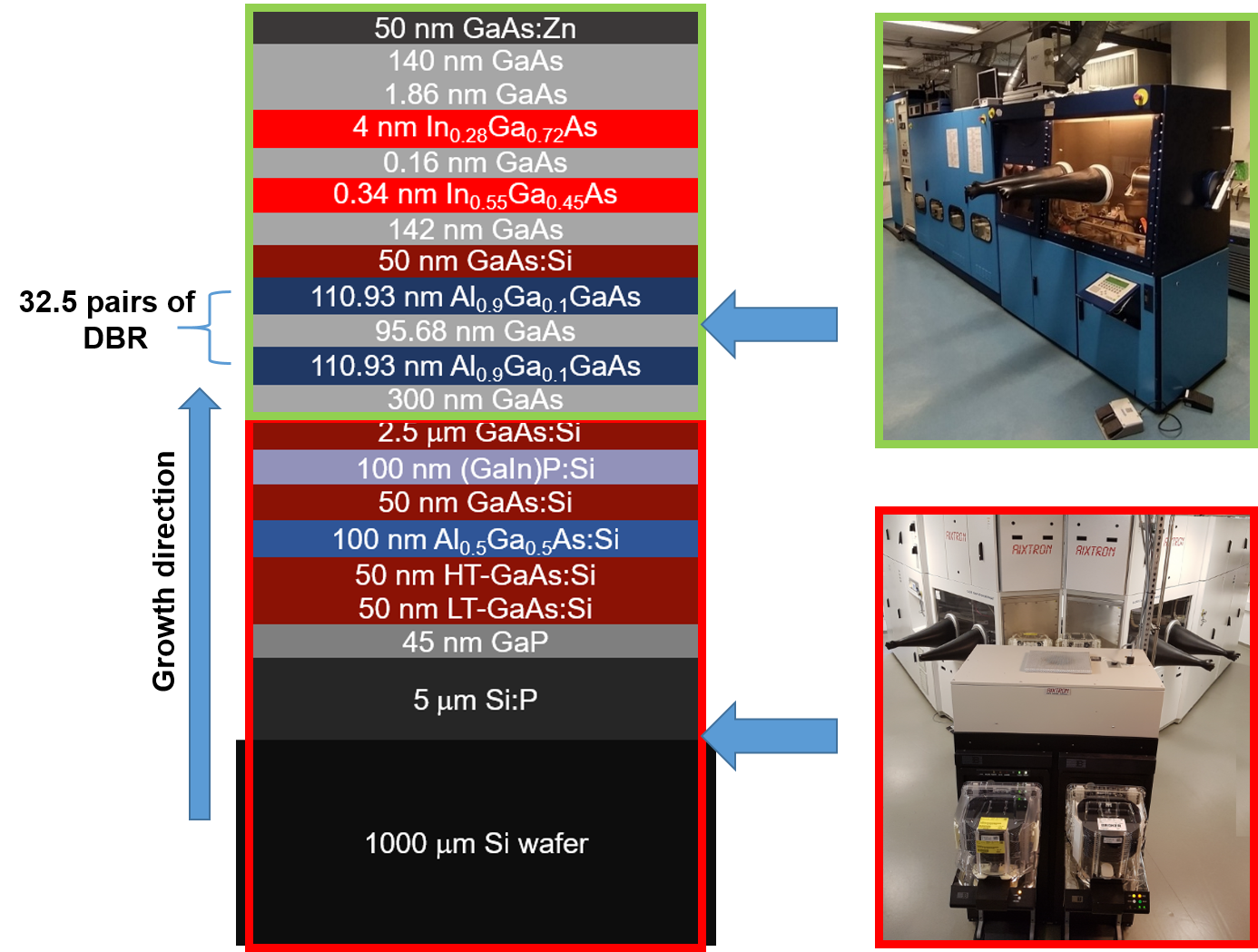}
  \caption{Heterostructure design and MOCVD growth setup for III-V quantum dot devices on silicon. (a) Schematic cross-section of the QD-heterostructure grown on a silicon substrate. (b) MOCVD reactor used for growth of the upper III-V layers, comprising the DBR, QDs, and cavity section. (c) MOCVD reactor used for growth III-V layers on silicon substrate, including GaP nucleation and defect filtration layers.}
  \label{Fig S 1}
\end{figure}

A schematic overview of the epitaxial layer structure is shown in Fig. S~\ref{Fig S 1}(a). The epitaxial template was realized on exact-oriented 300~mm Si (001) wafers ($\pm$0.5°) using a GaP-on-Si integration scheme. The first stage of the epitaxial growth was carried out in an AIXTRON Crius R CCS twin-reactor system, as illustrated in Fig. S~\ref{Fig S 1}(c). A 5~µm thick Si:P buffer layer was initially grown to provide a smooth and electrically conductive surface, followed by the growth of a thin GaP nucleation layer forming the III-V template on silicon \cite{VOLZ201137}. The use of GaP is critical for monolithic III-V integration on Si. Owing to its small lattice mismatch with silicon ($\sim$0.36\%). GaP enables comparatively low-defect heteroepitaxy and acts as a polarity transition layer, allowing the growth of polar III-V materials on the non-polar Si surface while suppressing the formation of anti-phase domains (APDs), which would otherwise degrade the structural quality of the epitaxial layers. Following GaP nucleation, a GaAs buffer layer was grown on the GaP/Si template. Owing to the $\sim$3.7\% lattice mismatch between GaAs and GaP, strain relaxation occurs through the formation of misfit dislocations at the GaAs/GaP interface, generating threading dislocations (TDs) that propagate toward the surface. To suppress their propagation, a sequence of AlGaAs/GaAs and GaInP dislocation filter layers (DFLs) was incorporated to promote the bending and annihilation of threading dislocations. Such strain-engineered layer stacks are widely used in III-V heteroepitaxy on silicon and effectively suppress the propagation of threading dislocations, resulting in a material quality sufficiently high to support narrow excitonic emission and maintain high single-photon purity in the telecom regime \cite{Limame2025}. The template growth was completed with a $\sim$2.5~µm thick GaAs:Si layer, forming a virtual GaAs substrate suitable for subsequent device growth.

The as-grown 300~mm wafer was then cleaved into 4~cm~$\times$~4~cm dies and transferred to a second MOCVD reactor AIXTRON 200/4 as illustrated in Fig. S~\ref{Fig S 1}(b), optimized for high-uniformity growth of optical and electronic III-V heterostructures. To establish a pristine growth template, the substrate was heated to 735~$^\circ$C to remove native surface oxides, followed by the growth of a 300~nm undoped GaAs layer at a high V/III ratio ($\approx 200$), resulting in an atomically smooth surface. A 33.5-pairs GaAs/$\mathrm{Al}_{0.9}\mathrm{Ga}_{0.1}\mathrm{As}$ DBR, comprising 95.68~nm GaAs and 110.93~nm $\mathrm{Al}_{0.9}\mathrm{Ga}_{0.1}\mathrm{As}$ quarter-wave layers, was then grown to provide a backside mirror with high reflectivity in the telecom O-band spectral range. Above the DBR, 192~nm of GaAs was grown to form the lower half of the $\lambda$-cavity. This region incorporates a 50~nm Si-doped GaAs layer ($n \approx 2 \times 10^{18}\,\mathrm{cm}^{-3}$), which serves as the n-type contact, followed by 142~nm of intrinsic GaAs to establish a field-free region around the active emitters. The substrate temperature was then reduced to 500~$^\circ$C for the growth of the QD layer. A 0.34~nm $\mathrm{In}_{0.55}\mathrm{Ga}_{0.45}\mathrm{As}$ wetting layer was grown, followed by a $\sim$60~s growth interruption to promote the nucleation of self-assembled QDs. The QDs were immediately capped with a thin GaAs layer to stabilize their morphology, before being overgrown with 4 nm of composition-graded InGaAs. This graded InGaAs layer functions as a strain-reducing layer (SRL), modifying the local strain environment, reducing hydrostatic compression, and redshifting the emission into the telecom O-band~\cite{Nishi1999, HOSPODKOVA2007}. In addition, the SRL improves dot uniformity and thermal stability by suppressing indium out-diffusion and limiting alloy intermixing during subsequent high-temperature processing~\cite{Grose2021}. Finally, the SRL was capped with 1.86~nm of GaAs, after which the substrate temperature was increased to 615~$^\circ$C to desorb residual indium. To complete the upper half of the $\lambda$-cavity, a 190~nm GaAs layer was grown at 615~$^\circ$C. The uppermost 50~nm of this region incorporates a Zn-doped GaAs layer ($p \approx 1.5 \times 10^{19}\,\mathrm{cm}^{-3}$), serving as the p-type contact and thereby completing the p-i-n structure, as shown in Fig. S~\ref{Fig S 1}(a).

\section{Device concept and numerical modeling}\label{sec2}
\begin{figure}[h]
 \centering
  \includegraphics[width=1\textwidth]{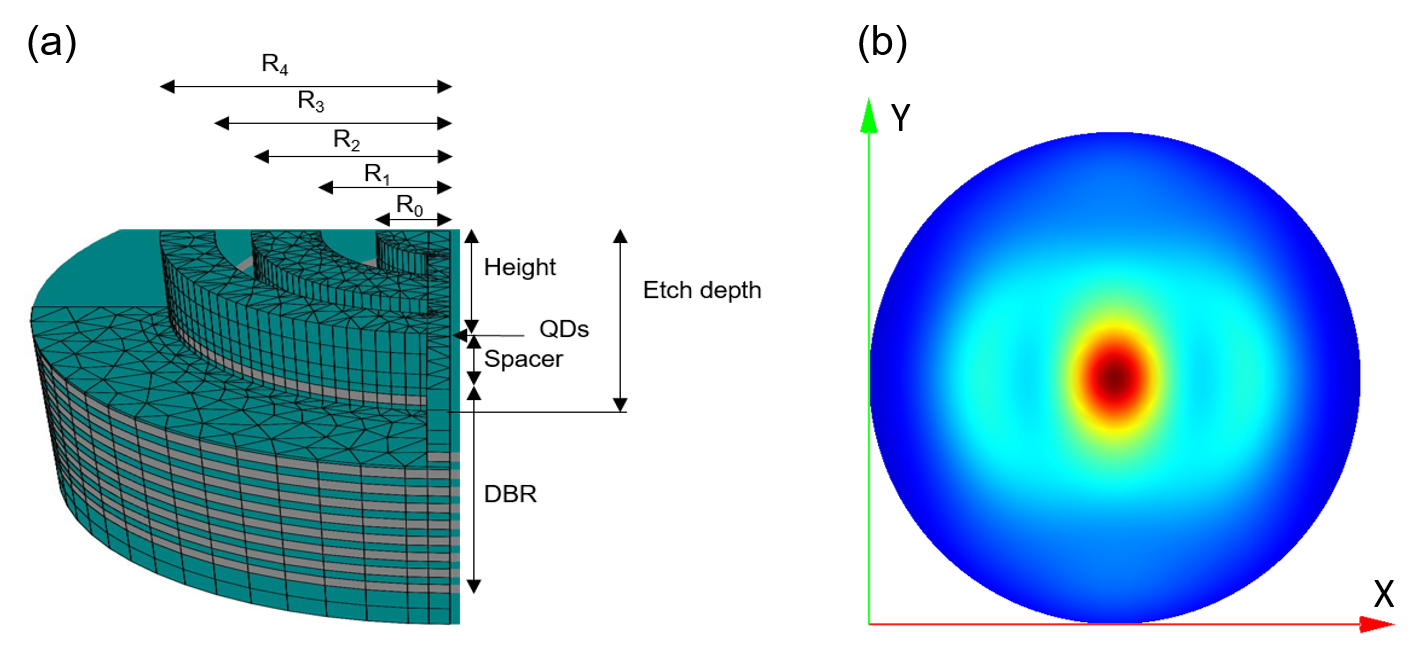}
  \caption{Design and simulated emission pattern of an eCBG QD resonator. (a) Three-dimensional schematic of the eCBG resonator, indicating key structural parameters and the QD positioned at the cavity center within the $\lambda$-cavity.
(b) Simulated far-field intensity distribution of the fundamental cavity mode, showing a directional, near-Gaussian radiation pattern consistent with efficient vertical photon extraction.}
  \label{fig S 2}
\end{figure}

Fig. S~\ref{fig S 2}(a) illustrates the three-dimensional schematic of the QD-eCBG resonator. The structure serves as the basis for full three-dimensional finite element method (FEM) simulations performed using the JCMsuite solver (JCMwave GmbH) to optimize the device geometry for the given epitaxial layer stack. The structure consists of a central mesa containing the QDs, positioned at the center of the $\lambda$-cavity to maximize coupling to the fundamental cavity mode. Surrounding the mesa, concentric annular trenches form a radial Bragg grating (CBG) that redirects in-plane guided modes toward vertical emission. The parameters $R_{0}$--$R_{4}$ denote the radii defining the central mesa and successive grating rings that establish the Bragg periodicity. In addition, a narrow ridge connects the central mesa to the outer contact region, providing the electrical conduction path necessary for operation of the vertical p-i-n diode structure.
The geometric configuration yielding the highest photon outcoupling into the collection optics (NA = 0.81) was identified by systematically varying key design parameters, including the grating period, ring thickness, etch depth, mesa diameter, and ridge width. The structure produces a highly directional far-field emission pattern for the fundamental cavity mode (Fig. S~\ref{fig S 2}(b)), exhibiting a near-Gaussian profile centered along the vertical axis. The resulting optimized device parameters are
summarized in Table~\ref{Table S 1}.

\begin{table}[ht]
 \caption{Optimized eCBG geometry parameters. Abbreviations: $Wl$ = design wavelength, $R_0$ = mesa radius, $R_1$–$R_4$ = Bragg ring radii, $n$ = number of rings, $w$ = ridge width, $d$ = etch depth, PEE = photon extraction efficiency.}
 
 \label{Table S 1}
  \setlength{\tabcolsep}{10.5pt}
 \begin{tabular}{ccccccccccc}
 \hline
 $Wl$ (nm) & $n$ & $R_0$ (nm) & $R_1$ (nm) & $R_2$ (nm) & $R_3$ (nm) & $R_4$ (nm) & $w$ (nm) & $d$ (nm) & PEE (\%) \\
\hline
 1303 & 2 & 275 & 1350 & 2203 & 3075 & 4102 & 150 & 660 &  31 \\
 \hline
 \end{tabular}
\end{table}

\section{Experimental setup}\label{sec3}

\begin{figure}[!h]
 \centering
  \includegraphics[width=1.0\textwidth]{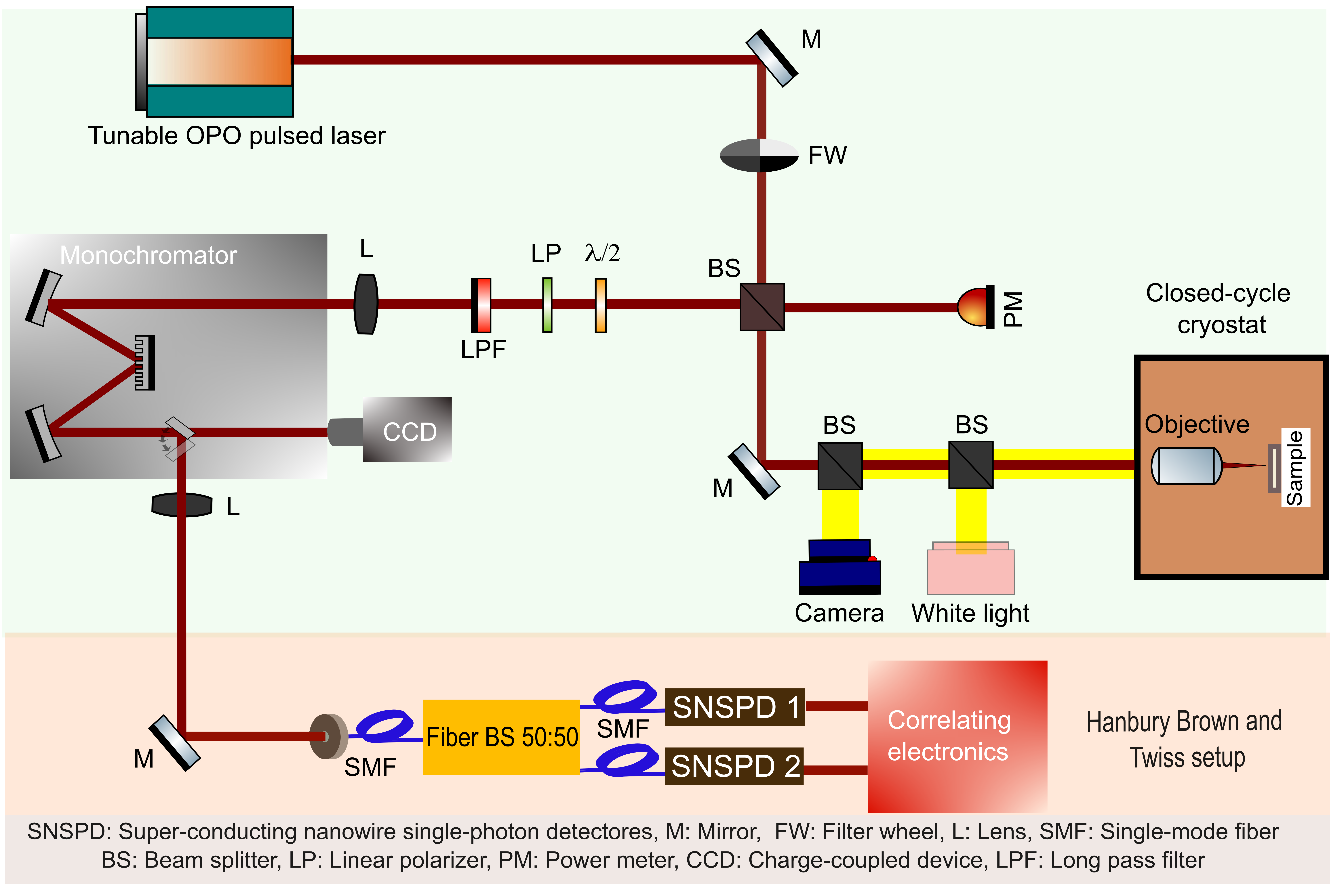}
  \caption{Schematic representation of the micro-photoluminescence ($\mu$PL) setup used for optical and quantum optical studies of QD-eCBGs.} 
  \label{fig S 3}
\end{figure}

Optical characterization was performed using a $\mu$PL setup comprising a closed-cycle cryostat (Fig. S~\ref{fig S 3}). The cryostat is equipped with a three-axis piezoelectric nanopositioning stage for precise alignment of the device relative to the excitation beam and electrical feedthroughs connected to the mounted chip carrier, allowing application of an external low-noise voltage source for controlled biasing of the vertical p-i-n diode during optical and quantum-optical measurements. Excitation and collection were carried out through a high-numerical-aperture objective lens (NA = 0.81), providing efficient light coupling and diffraction-limited spatial resolution. Pulsed excitation was supplied by an 80~MHz optical parametric oscillator (OPO) delivering wavelength-tunable laser pulses. The excitation beam was directed onto the sample via a 10:90 beam splitter (excitation:detection ratio), allowing excitation and collection through a common optical path. The emitted photoluminescence was spectrally dispersed using a monochromator equipped with interchangeable 150~lines/mm and 900~lines/mm gratings, corresponding to spectral resolutions of approximately 150~$\mu$eV and 20~$\mu$eV, respectively. Spectral detection was performed using an InGaAs line-array detector. For second-order photon autocorrelation measurements, the monochromator output was redirected to a Hanbury Brown and Twiss (HBT) configuration. The emission was coupled into a single-mode fiber beam splitter and routed to two superconducting nanowire single-photon detectors (SNSPDs) with a timing resolution of approximately 30~ps. Coincidence events were recorded using time-correlated single-photon counting (TCSPC) electronics, from which the second-order autocorrelation function $g^{(2)}(\tau)$ was extracted.

\section{Current–voltage (I-V) characteristics}\label{sec4}

\begin{figure}[!h]
 \centering
  \includegraphics[width=0.6\textwidth]{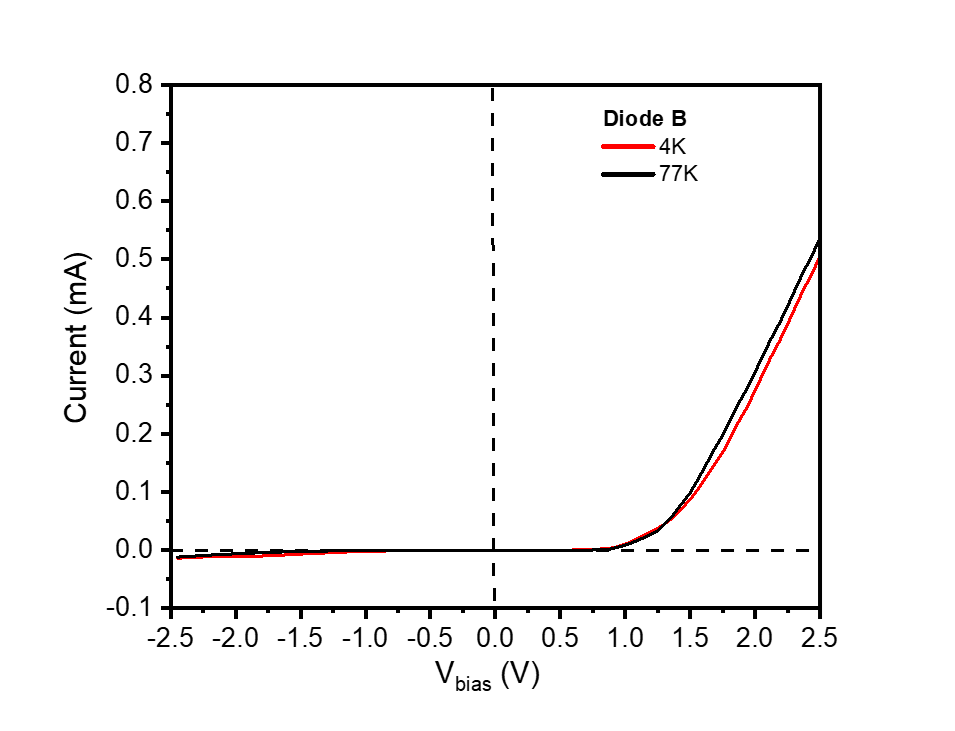}
  \caption{Cryogenic I-V characteristics of the p-i-n diode at 4 K and 77 K }
  \label{fig S 4}
\end{figure}

Current-voltage (I-V) measurements were performed at 4~K and 77~K to assess the electrical characteristics of the fabricated p-i-n diode, Fig. S~\ref{fig S 4}. The measurements were carried out on one of the four fabricated diodes (diode B), which was wire-bonded and contained the integrated QD-eCBGs. The devices exhibit clear rectifying behavior at both temperatures, with low reverse leakage current (around 8 $\mu$A) and well-defined forward conduction, confirming proper diode operation after contact fabrication and wire bonding. The preserved diode characteristics confirm stable electrical operation under cryogenic conditions and enable controlled modulation of the internal electric field for Stark tuning.

\section{Bias-dependent optical and quantum-optical measurement of eCBG$_{1}$}\label{sec4}

\begin{figure}[h]
 \centering
  \includegraphics[width=1.0\textwidth]{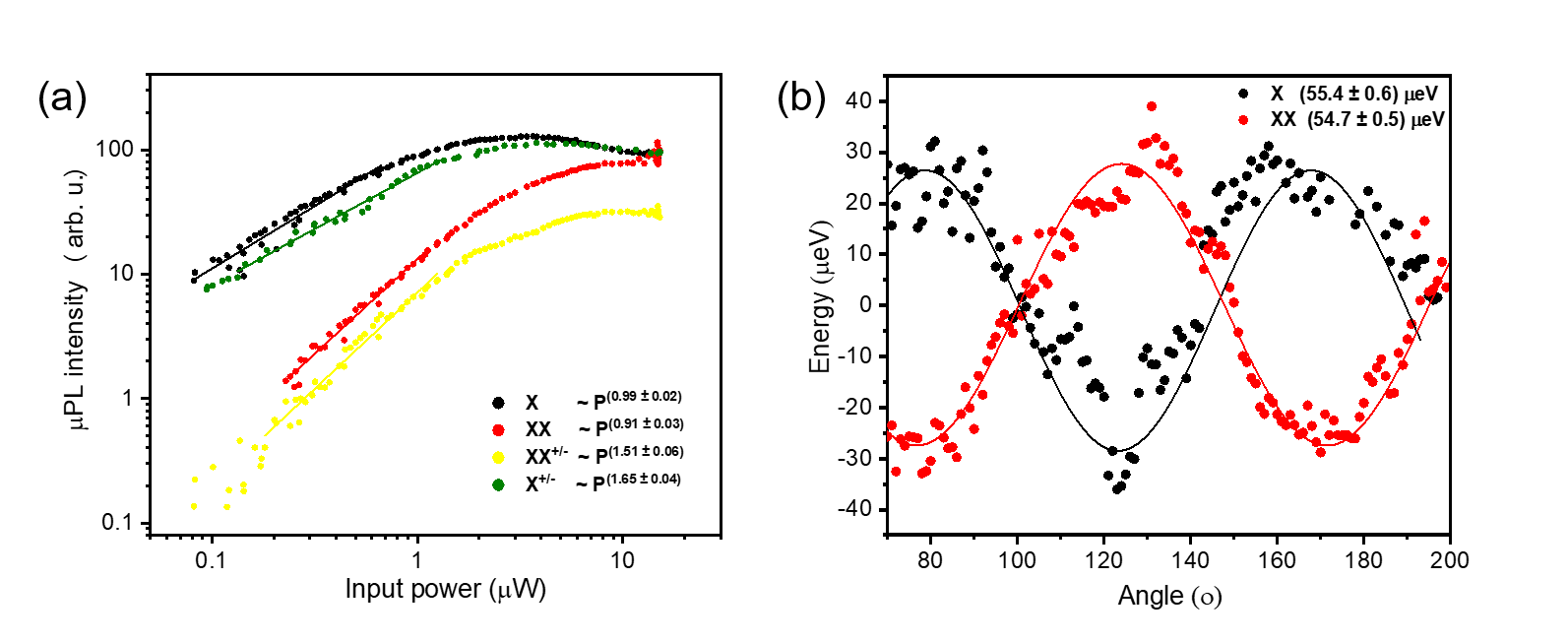}
  \caption{Excitation-power dependence and polarization-resolved spectroscopy of excitonic transitions of eCBG$_1$. (a) Double-logarithmic plot of the $\mu$PL intensity as a function of excitation power, showing the power dependence excitonic transitions: neutral exciton (X, black), charged exciton (X$^{\pm}$, green), biexciton (XX, red), and charged biexciton (XX$^{\pm}$, yellow). Solid lines represent linear fits in log–log scale. (b) Polarization-resolved $\mu$PL measurements of the same transitions, revealing a fine-structure splitting.
}
  \label{fig S 5}
\end{figure}

Power-dependent $\mu$PL measurements were performed by varying the excitation power and integrating the spectrally filtered emission lines to identify the excitonic complexes of eCBG$_{1}$. The neutral and charged exciton transitions show nearly linear scaling, whereas the biexciton-related lines exhibit superlinear behavior consistent with multi-carrier recombination Fig. S~\ref{fig S 5}(a). For polarization-resolved measurements, a half-wave plate and linear polarizer were introduced in the detection path, revealing a sinusoidal modulation of the neutral exciton energy with polarization angle Fig. S~\ref{fig S 5}(b), confirming its fine-structure splitting of $\Delta_{\mathrm{FSS}} = (55.5 \pm 0.6)\mu$eV. Detailed information about the excitonic line can be found in Table~\ref {Table S 2}. 

\begin{table}[h]
 \caption{Emission lines observed from eCBG$_{1}$ and their relevant parameters, Abbreviations: $\lambda$ = emission wavelength, $E$ = emission energy, $m$ = power exponents, $\Delta E$ = energy deviations, $LW$ = emission linewidth.} 
 \label{Table S 2}
 \setlength{\tabcolsep}{27pt}
 \begin{tabular}{ccccccccccc}
 \hline
 emission & $\lambda$ (nm) & $E$ (meV) & $m$  & $\Delta E$ ($\mu$eV)  &  LW ($\mu$eV) \\
\hline
X & 1302.83 & 951.65 & 0.99 & 55.40 & 170.93  \\
 \hline
 X$^{\pm}$ & 1304.73 & 950.27 & 0.91 & none & 138.66   \\
 \hline
 XX & 1307.05 & 948.58 & 1.51 &  54.75 & 130.30  \\
 \hline
 XX$^{\pm}$ & 1308.08 & 947.83 & 1.65 & none & 155.73 \\
 \hline
 \end{tabular}
\end{table}

\begin{figure}[!h]
 \centering
  \includegraphics[width=1.0\textwidth]{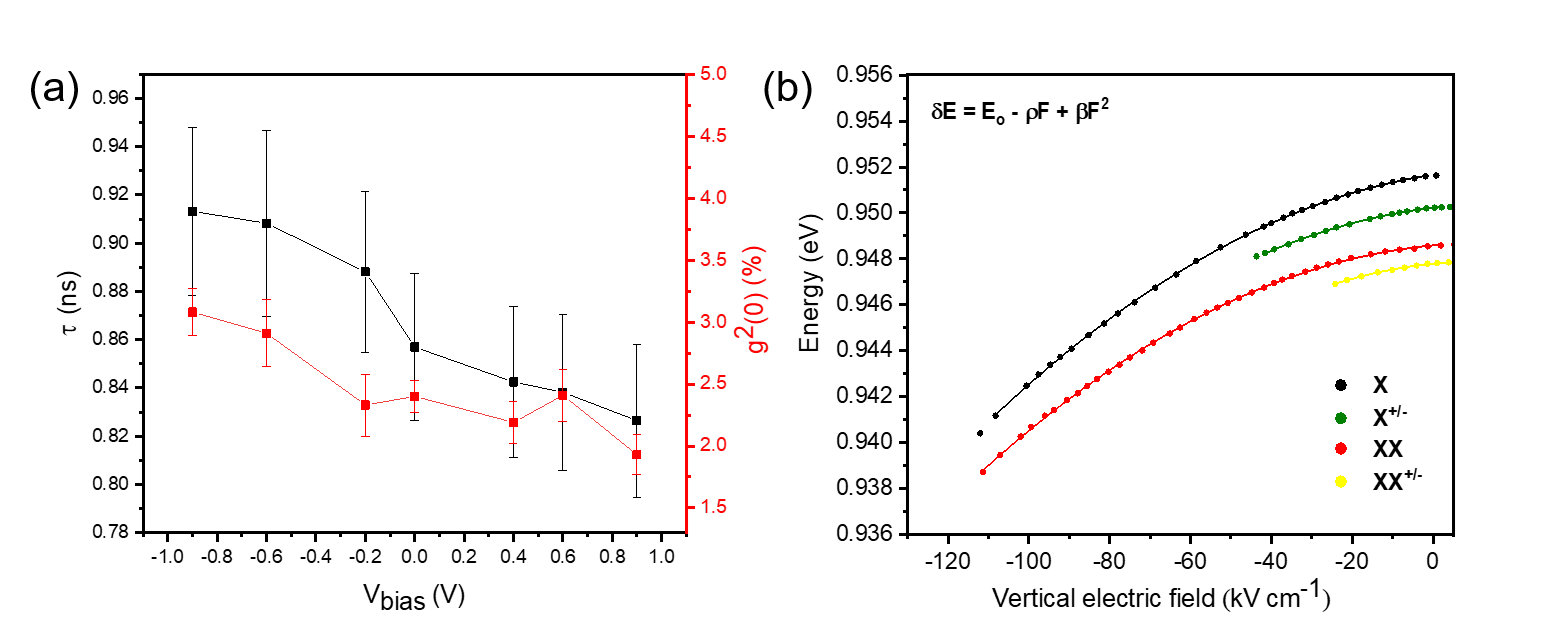}
  \caption{Voltage-dependent optical characterization and Stark-effect analysis of eCBG$_1$. (a) Voltage-dependent optical and quantum-optical characterization of eCBG$_{1}$. The radiative lifetime and raw $g^{(2)}(0)$ values are shown as a function of applied bias. (b) QCSE analysis of multiple excitonic transitions. The field-dependent emission energies are fitted using the quadratic Stark model to extract the permanent dipole moments and polarizabilities of the respective excitonic states. The solid lines denote the model fits.
}
  \label{fig S 6}
\end{figure}

Figure S~\ref{fig S 6}(a) shows the bias-dependent radiative lifetime of the neutral exciton together with the corresponding raw $g^{(2)}(0)$ values for device eCBG$_1$. With increasing reverse bias, the exciton lifetime exhibits a gradual increase, consistent with the field-induced reduction of the electron-hole wavefunction overlap associated with the quantum-confined Stark effect (QCSE)\cite{Mille1984, Bennett2010}. The consistently low raw $g^{(2)}(0)$ values ($<3\%$) at saturation indicate that electrical Stark tuning does not lead to a pronounced degradation of the single-photon purity over the investigated bias range.

The bias-dependent emission energies were analyzed within the framework of the QCSE, which describes the modification of confined excitonic states in semiconductor nanostructures under an applied electric field. In quantum dots, the strong three-dimensional confinement leads to discrete electron and hole states whose energies and spatial distributions can be perturbed by an external electric field. When a vertical electric field $F$ is applied across the heterostructure, the excitonic transition energy shifts due to the field-induced displacement and polarization of the electron and hole wavefunctions along the growth direction, which modifies their confinement energies and mutual Coulomb interaction \cite{Fry_2000, Barker_2000}.

 In the presence of a vertical electric field $F$, the excitonic transition energy can be expressed to second order as\begin{equation} E(F) = E_0 - pF + \beta F^2,\end{equation}   \cite{Finley_2004, Mar_2017} where $E_0$ denotes the zero-field transition energy, $p$ is the built-in dipole moment, and $\beta$ represents the polarizability.

From this model, we extract the electron-hole separation $p/e$ ($nm$)  and the polarizability $\beta$ $(\mu\mathrm{eV}\,/(\mathrm{kV}\,/\mathrm{cm})^{2})$ for different excitonic transitions.
The values obtained are as follows: for neutral exciton (X, black) $p/e = -(0.276 \pm 0.006)$ and $\beta = -(0.637 \pm 0.005)$; for charged exciton (X$^{\pm}$, green) $p/e = -(0.206 \pm 0.007)$ and $\beta = -(0.651 \pm 0.017)$; for biexciton (XX, red) $p/e = -(0.265 \pm 0.019)$ and $\beta = -(0.564 \pm 0.014)$; for charged biexciton (XX$^{\pm}$, yellow) $p/e = -(0.183 \pm 0.013)$ and $\beta = -(0.725 \pm 0.072)$. The extracted electron–hole separations lie in the range of $\sim0.18$–$0.28$ nm, which is consistent with previously reported values for In(Ga)As quantum dots obtained from Stark spectroscopy \cite{Fry_2000, Findeis_2001, Oulton_2002, Petruzzella_2015}. A reduction of $|p/e|$ is observed for the charged complexes compared to the neutral exciton, consistent with earlier studies showing that additional carriers modify the Coulomb interaction and partially redistribute the electron–hole wavefunctions \cite{Finley_2004, Mar_2017}.

\section{Temperature dependent optical measurement of eCBG$_{1}$ }\label{sec5}

\begin{figure}[!h]
 \centering
  \includegraphics[width=1\textwidth]{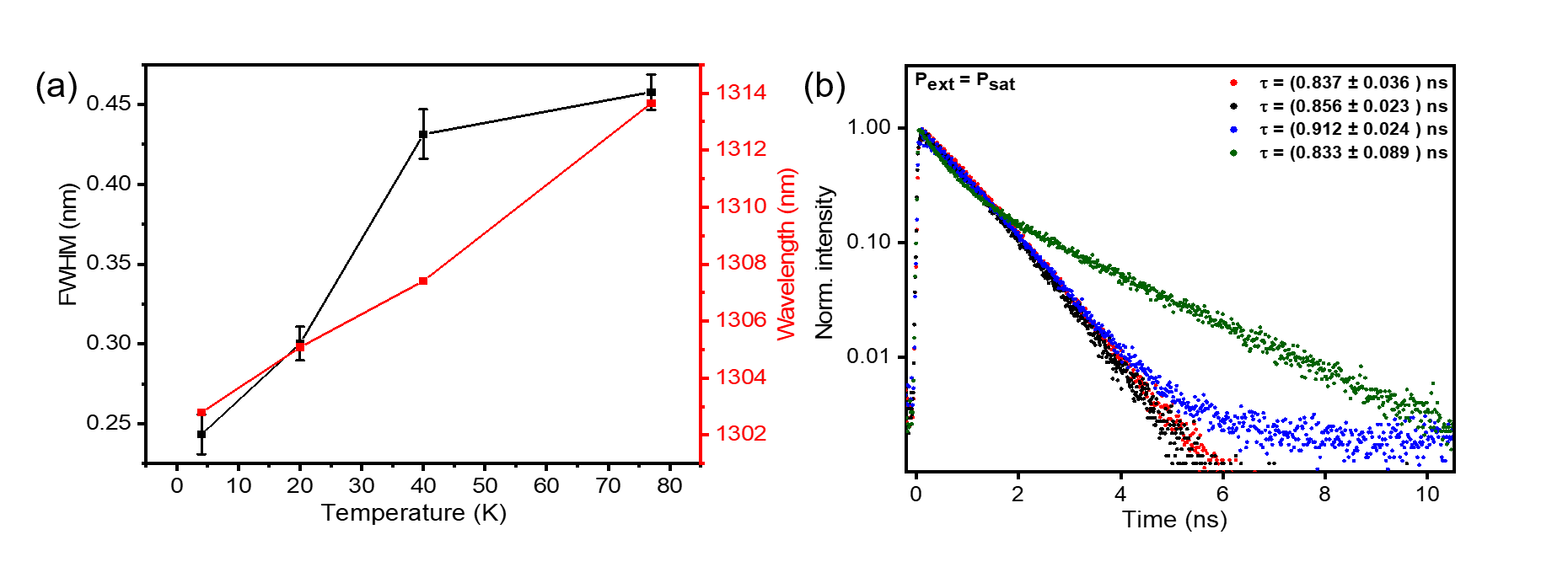}
  \caption{Temperature-dependent spectral and radiative properties of the excitonic transition of eCBG$_1$. (a) The FWHM (black data points) and emission wavelength (red data points) of the excitonic (X) transition as a function of temperature. (b) Temperature-dependent decay time of X. An exponential fit is applied to extract the lifetime across the temperature series. }
  \label{fig S 7}
\end{figure}

Figure S~\ref{fig S 7}(a) shows the temperature dependence of the neutral exciton (X) emission wavelength and linewidth for device eCBG$_1$. Consistent with the trends discussed in the main text, the emission exhibits a monotonic redshift (from 1302.79 nm to 1313.65 nm) accompanied by a progressive linewidth broadening (from 0.24 nm to 0.46 nm) with increasing temperature (from 4~K to 77~K)~\cite{Bayer2002, Holewa2020TelecomQD}.

Figure S~\ref{fig S 7}(b) presents the temperature-dependent time-resolved $\mu$PL measurements of the neutral exciton (X) emission from device eCBG$_1$. The decay curves were recorded under saturation excitation and fitted with a single-exponential function to extract the exciton lifetime at each temperature. The extracted lifetimes remain nearly constant across the investigated temperature range, with values of $\tau = (0.837 \pm 0.036)\,\mathrm{ns}$ at 4~K, $(0.856 \pm 0.023)\,\mathrm{ns}$ at 20~K, $(0.912 \pm 0.024)\,\mathrm{ns}$ at 40~K, and $(0.833 \pm 0.089)\,\mathrm{ns}$ at 77~K, in good agreement with typical values reported for self-assembled InGaAs/GaAs QDs \cite{Grose2021}. This weak temperature dependence indicates that the intrinsic radiative recombination dynamics of the exciton remain largely unaffected by thermal activation up to 77~K, confirming robust carrier confinement in the telecom InGaAs QDs  \cite{Carmesin_2018}. A bi-exponential decay occurs at 77 K, with a short component having a decay time of $\tau = (0.833 \pm 0.089)\,\mathrm{ns}$ and a long component having a decay time of $\tau = (2.069 \pm 1.228)\,\mathrm{ns}$. This behavior is attributed to thermally activated carrier redistribution involving excited QD states or the wetting layer \cite{Dalgarno2008}. In this regime, phonon-assisted scattering and thermal population of excited states can temporarily depopulate the excitonic ground state. Subsequent carrier recapture repopulates the ground state and results in delayed emission, introducing a slow component in the decay dynamics while leaving the intrinsic radiative lifetime of the exciton largely unaffected.
